\documentclass[preprint,showpacs,
superscriptaddress]{revtex4}
\usepackage[latin1]{inputenc}
\usepackage[T1]{fontenc}
\usepackage[english]{babel}
\usepackage{exscale,times}
\usepackage{amsmath,amssymb}
\usepackage{graphicx}
\usepackage{setspace}

\def\Nabla{\boldsymbol{\nabla}}

\def\Div{\boldsymbol{\nabla\cdot}}
\def\We{{\rm We}}
\def\B{\Gamma}
\def\mueffi{\mu^{\rm eff}}
\def\mueffitil{{\tilde \mu}^{\rm eff}}
\def\mueff{\mu_{\rm eff}}
\def\muefftil{{\tilde\mu}_{\rm eff}}

\def\ratio{\nu}
\def\zet{\muefftil}

\def\vv{\vec{v}}

\begin{document}

\title{Phase-field simulations of viscous fingering in shear-thinning fluids}

\author{S{é}bastien Nguyen}
\affiliation{PPMD, ESPCI, CNRS, 10 rue Vauquelin, 75005 PARIS, France}
\affiliation{PMC, Ecole Polytechnique, CNRS, rte de Saclay , 91128 PALAISEAU, France}

\author{R. Folch}
\affiliation{Departament d'Enginyeria Qu\'{\i}mica, 
Universitat Rovira i Virgili, Av. dels Pa\"{\i}sos Catalans, 26, 
E-43007 Tarragona, Spain}

\author{Vijay K. Verma}
\affiliation{Department on Chemical Engineering, Indian Institute of Technology Guwahati,
Guwahati, Assam 781039}
\author{Herv{é} Henry}\affiliation{PMC, Ecole Polytechnique, CNRS, rte de Saclay , 91128 PALAISEAU, France}

 \author{Mathis Plapp}
\affiliation{PMC, Ecole Polytechnique, CNRS, rte de Saclay , 91128 PALAISEAU, France}

\begin{abstract}
A phase-field model for the Hele-Shaw flow of non-Newtonian fluids
is developed. 
It extends a previous
model for
Newtonian fluids 
to a wide range of shear-dependent fluids.
The model is applied to perform simulations of viscous fingering in shear-
thinning fluids, and it is found to be capable of describing the complete crossover 
from the Newtonian regime at low shear rate to the strongly shear-thinning 
regime at high shear rate.
The width selection of a single steady-state finger is studied
in detail for a 2-plateaux shear-thinning law (Carreau law) in both its
weakly and strongly shear-thinning limits, and the results 
are related to previous analyses.
In the strongly shear-thinning regime 
a rescaling is found for power-law (Ostwald-de-Waehle) fluids that
allows for a direct comparison between
simulations and experiments without any adjustable parameters,
and good agreement is obtained.

\end{abstract}

\date{\today}

\maketitle

\section{Introduction}

The Saffman-Taylor instability occurs when a fluid is pushed by
another one of lower viscosity in a confined geometry, such as
porous media or a Hele-Shaw cell. It leads to the emergence
of complex interfacial patterns whose shape is reminiscent of fingers. 
The study of this phenomenon \cite{Saffman58,McLean81} has 
helped to establish much of our current knowledge 
on the self-organization of branched 
patterns \cite{Paterson81,Bensimon86,Kessler88}. Indeed, 
viscous fingering can be studied under well-controlled conditions 
in the laboratory using Hele-Shaw cells, where the flow is 
confined to a narrow gap between two parallel plates. In this geometry, 
the full flow can be well descibed by an effective two-dimensional problem,
which greatly simplifies both theoretical analysis and numerical
simulations. 

For two Newtonian fluids of strongly different viscosities, 
our understanding is fairly complete. In a channel geometry,
the instability of a flat interface and
the subsequent evolution results in the formation of a single
finger, the so-called Saffman-Taylor finger \cite{Saffman58}.
Its relative width with respect to the channel 
is selected by a subtle interplay between viscous dissipation 
and the surface tension of the interface, which acts as a singular
perturbation. In a radial geometry,
where the low-viscosity fluid is injected through a central
inlet, fingers are not stable and exhibit repeated tip-splitting
to form highly ramified patterns \cite{Paterson81}.

Much less is known about viscous fingering in non-Newtonian fluids. 
Numerous experiments have revealed
that a wide variety of patterns can be formed,  including 
finger patterns  close to the ones found in Newtonian
fluids with either narrowing or widening of the fingers,  
straight fingers in a radial geometry 
that do not exhibit tip splitting,
and  patterns that form angular branches and sharp tips,
reminiscent of crack networks (for a review,
see \cite{McCloud95}). 

It is clear that the selection of these patterns is governed by 
the nonlinearity of the fluid itself. More precisely, there is
a complex interplay between the geometry of the finger, which
determines the local flow pattern. The latter, in turn, modifies
the properties of the fluids. In the particular case of 
shear-thinning fluids, the dependency of the fluid viscosity on 
the local shear rate, which strongly varies in the vicinity 
of a finger tip, can create an effect which is akin to an 
interfacial anisotropy. The latter is known both from 
experiments \cite{Ben-Jacob85,Couder86,Rabaud88} and
theory \cite{Kessler86,Dorsey87} to profoundly affect
pattern selection. Its presence suppresses tip-splitting 
and favors the emergence of dendritic patterns with sidebranches. 
The transition from branching fingers to dendrites observed
in liquid crystals \cite{Buka87} can thus be explained, at
least qualitatively \cite{Folch00,Folch01}.
Furthermore, it is not surprising to see crack-like patterns
in viscoelastic fluids \cite{Lemaire91}, since a high shear around
the tip pushes the fluid into the elastic regime.

For a more detailed and quantitative investigation of this
relation between morphologies and the rheological properties of 
non-Newtonian fluids, precise numerical models would be very 
helpful. However, in mathematical terms, viscous fingering is normally 
formulated as a free boundary problem, which is quite difficult 
to handle numerically \cite{Tryggvason83,Hou94,Fast01,Fast04}.
To our best knowledge, simulations of non-Newtonian
viscous fingering using such methods
have remained limited to the case of shear-thinning fluids
in the weakly shear-thinning limit \cite{Fast01,Fast04}.
To overcome  the difficulties due to moving interfaces, diffuse-interface 
and phase-field methods have become popular in 
many different fields
\cite{Anderson98,Boettinger02,Chen02,Gonzalez-Cinca04,Karma01b}.
In phase-field models, a continuous scalar field, the phase 
field, is introduced to distinguish between the two domains 
occupied by the two fluids. All properties of the fluids
are interpolated through the diffuse interface, and the
motion of the phase field is coupled to the equations of
fluid dynamics. The original free boundary problem is
obtained in the limit of vanishing interface thickness.
While this approach  introduces an additional scale (the
interface thickness) into the problem, it removes the difficulties due to
explicit interface tracking (non-uniform length change of the
interfqce, topological changes).
Therefore, its implementation is straightforward.

In this paper, we develop a phase-field model
for Hele-Shaw flow in a wide class of fluids with a 
shear-dependent viscosity, by combining a phase-field
model for Newtonian viscous fingering previously developed 
by one of us \cite{Folch99a,Folch99b} with a rigorous procedure
for obtaining a generalized Darcy's law for non-Newtonian fluids
developed by Fast {\em et al.} \cite{Fast01}. The model is
implemented using a finite-difference scheme 
in conjunction with a standard SOR solver for the pressure
equation. We validate our model and implementation by a 
detailed comparison of the Newtonian case to the known 
sharp-interface solution. This  allows us to estimate 
the errors that are due to the finite interface thickness 
and the discretization. 

Although our model is capable of describing two non-Newtonian
fluids with general shear-dependent viscosity laws, we limit ourselves
to shear-thinning fluids  pushed
by a Newtonian fluid. Indeed, this is the setting where 
the most precise knowledge on pattern selection in 
non-Newtonian fluids is already available, and therefore 
it constitutes an excellent testing ground for our model. 
Data on the shape and width of steady-state fingers 
for shear-thinning fluids with a well-characterized viscosity 
law have been published \cite{Lindner00,Lindner02}. Furthermore, 
these data are in good agreement with theoretical studies that
predict a narrowing of the steady-state fingers with respect
to the Newtonian case \cite{Corvera98,BenAmar99}.

We perform simulations for two different viscosity laws, namely,
a two-plateau law used in the simulations of Refs.~\cite{Fast01,Fast04},
and the one-plateau law  which describes well, for the experimental flow regime,  
the fluids used in 
experiments of Refs.~\cite{Lindner00,Lindner02}. We study the
effect of the shear thinning on the selection of the finger
width, and demonstrate that our model is able to cover the
complete crossover from Newtonian behavior at low speed to
strong shear-thinning at high speeds. More precisely, the
selection of the finger width can be understood in terms
of two dimensionless parameters: the Weissenberg number $\We$,
which characterizes the strength of the shear-thinning effect,
and a dimensionless surface tension $\B$. In general, the finger 
width depends on both parameters. However, it turns out that
in the regime covered by the experiments \cite{Lindner00,Lindner02}, 
the viscosity law can be well described by a simple power law. In 
this case, the finger width depends only on a single parameter, 
which is a function of $\We$, $\B$ and the exponent of the 
viscosity law. In this regime, our simulations are in good agreement
with the experimental data of Refs.~\cite{Lindner00,Lindner02},
which demonstrates the capability of our model to yield
quantitatively accurate results.

The remainder of the paper is organized as follows:
Section~\ref{model} presents the theoretical framework,
the model and briefly discuss its numerical implementation. Results
are then presented in Sec.~\ref{results}, followed by
conclusions and perspectives in Sec.~\ref{conclusions}.

\section{Model}
\label{model}

\subsection{Sharp-interface equations}
\label{fbp}

We consider two incompressible, immiscible fluids (labeled 1 and 2) 
in a Hele-Shaw cell of width $W$ ($x$-direction), length $L$ 
($y$-direction) and gap $b$ ($z$-direction, $b\ll W<L$). The less 
viscous fluid 2 is injected at one end of the cell with a fixed 
flow rate $Q$, causing outflow of fluid 1 at the other end of the
cell with a velocity $U_\infty=Q/(bW)$. The interface between the
two fluids has a positive surface tension $\sigma$. Both fluids, 
may have a non-Newtonian shear viscosity that depends on the
local shear rate ${\dot\gamma}$, $\mu_i(\tau_i{\dot\gamma})$,
where $\tau_i$ is a characteristic relaxation time of fluid $i$;
we furthermore suppose that both viscosity laws have well-defined
Newtonian limits when $\dot\gamma\to 0$, which we will denote 
by $\mu_i^0$.

As usual in a Hele-Shaw cell at low velocities (where inertia can be
neglected), the scale separation between the
gap and the channel width makes it possible to simplify the full
three-dimensional flow problem by a long-wave approximation.
The resulting two-dimensional problem is stated, for each fluid,
in terms of the pressure field $p_i$ (which is constant across the 
gap) and the gap-averaged in-plane velocity $\vec u_i$. These
two-dimensional velocity fields remain incompressible,
\begin{equation}
\label{2incompressibilities}
\vec\nabla\cdot \vec u_i = 0, \qquad i=1,2.
\end{equation}
Furthermore, for Newtonian fluids, the local averaged velocity is 
proportional to the local in-plane pressure gradient, a relationship 
known as Darcy's law.
For non-Newtonian fluids, the relationship between $\vec u_i$ and
$\vec\nabla p_i$ becomes non-linear, but can formally still be written
as a generalized Darcy's law,
\begin{equation}
\label{2darcies}
\vec u_i = -\frac{b^2}{12\mueffi_i(b\tau_i|\vec\nabla p_i|/\mu_i^0)}
\vec\nabla p_i,\qquad i=1,2
\end{equation}
where $\mueffi_i$ is an {\em effective} viscosity, which can 
be related to the original shear-dependent viscosity
$\mu_i(\tau_i{\dot\gamma})$
for a large class of non-Newtonian fluids following the procedure
developed by Fast {\em et al.} \cite{Fast01}, which is summarized
and presented using  the notations of the present work in
Appendix~\ref{sec_darcy}. We have included the constants $b$,
$\tau_i$ and $\mu_i^0$ in the argument of the effective viscosity
to emphasize  that this argument  is indeed a dimensionless shear. 
The characteristic local shear rate can be estimated by 
the ratio of the gap-averaged velocity and the cell gap $b$; the 
order of magnitude of the velocity, in turn, is given 
by $|\vec\nabla p|/\mu_i^0$ (see Appendix \ref{sec_darcy} for details).
Note that we have chosen to express the viscosity as a function of 
the pressure gradient (and not of the velocity as in
\cite{Corvera98,BenAmar99,Lindner00,Lindner02}) in order to 
formulate the model  in terms of the interface geometry and the pressure 
field only. One should note that for a vanishing  shear rate ($|\vec\nabla 
p_i|\to 0$), we have
$\mueffi_i\to \mu_i^0$, and Eq.~(\ref{2darcies}) reduces to the 
standard Darcy's law.

 Since we are considering two fluid regions separated by an interface, we
have to specify the boundary conditions at the interface:
 \begin{eqnarray}
\label{laplace}
p_2-p_1&=&\sigma\kappa,\\
\label{impenetrability}
\hat r\cdot \vec u_1 &=&\hat r\cdot \vec u_2 =v_n,
 \end{eqnarray}
where $\kappa$ is the  interface curvature (in the plane of flow), $\sigma$ is 
the surface tension and
$\hat r$ is the unit vector normal to the interface pointing into fluid 1. 
Equation (\ref{laplace}) is simply the Laplace law, where the curvature
of the meniscus between the plates has been omitted under the assumption
that it is constant. Eq. \ref{impenetrability} simply assures the
impenetrability of the two fluids. 

In order to make this formulation more directly amenable to 
the construction of a phase-field model, we rewrite the above
equations in terms of a single set of fields and material 
properties \cite{Bedeaux76},
\begin{eqnarray}
p & = & \chi_1p_1+\chi_2p_2,
\label{begin_chi} \\
\vec u & = & \chi_1\vec u_1+\chi_2\vec u_2, \\
\label{viscosity}
\mueff & = & \chi_1\mueffi_1(\frac{b\tau_1|\vec\nabla p|}{\mu_1^0})+
\chi_2 \mueffi_2(\frac{b\tau_2|\vec\nabla p|}{\mu_2^0}),
\end{eqnarray}
where $\chi_1(\vec x)$ and $\chi_2(\vec x)$ are the characteristic
functions of the domains occupied by the two fluids (that is,
$\chi_i(\vec x)=1$ if the point $\vec x$ is occupied by fluid $i$,
and $0$ otherwise). We thus reduce
Eqs.~(\ref{2darcies}, \ref{laplace} and \ref{2incompressibilities})
to just two:
\begin{equation}
\label{darcy}
\vec u = -\frac{b^2}{12\mueff}
\left [ \vec\nabla p + \sigma\kappa\delta_\Sigma\hat r \right ],
\end{equation}
\begin{equation}
\label{incompressibility}
\vec\nabla\cdot \vec u = 0,
\end{equation}
where $\delta_\Sigma$ is a surface delta function (that is, a Dirac
delta function located on the sharp interface $\Sigma$ separating
the two fluid domains \cite{Bedeaux76}).
Now  all fields, material properties and equations must 
be understood in the sense of mathematical distributions.
As such, these equations, apart from their obvious limits
at each side of the interface, are to be understood when integrated 
across the interface. In particular, integrating the normal
projection of the velocity  times the effective
viscosity in Eq.~(\ref{darcy})
across the interface gives  the Laplace pressure drop
of Equation (\ref{laplace}).
Similarly, the condition of zero divergence of Equation
(\ref{incompressibility})
relates the normal and tangential components of the fluid velocity.
The condition of incompressibility, when applied on
the very interface, translates into impenetrability of the two fluids.

\subsection{Phase-field model}
\label{pfm}

In this section, we present the phase field approach to this problem. We first
give a brief description of the phase field (denoted by $\phi$) and show how
using it instead of the indicator functions, the flow equations
(\ref{darcy}) and (\ref{incompressibility}) are modified. 
Then we present the evolution equation for the phase field and give a rationale 
for its construction. Finally, we comment briefly on how the phase-field
model is an approximation of the sharp-interface model.

The idea underlying the phase-field model is to introduce an additional field
($\phi$) that indicates in which phase (here, in which fluid) the system is at
a given space point. For the sake of simplicity and without any loss of 
generality, we consider that in fluid 1 (resp. 2) , $\phi=1,\mbox{ (resp.
}-1)$. In addition, when crossing the interface the phase field exhibits
a smooth front (kink) of finite width. In this general framework, 
the indicator functions and the $\delta_\Sigma$ function are approximated by
 \begin{eqnarray}
  \chi_1 &\to& (1+\phi)/2,\\ 
  \chi_2 &\to& (1-\phi)/2,\\
  \delta_\Sigma &\to& |\nabla \phi|/2.
 \end{eqnarray}   
  
Then, replacing $\chi_1$ and $\chi_2$ by their smoothed expressions,
the effective viscosity of Eq.~(\ref{viscosity}) becomes
   \begin{equation}\label{interpolatedviscosity}
\mu_{\mbox{eff}}(\phi)=\frac{1+\phi}{2}\mu^{\mbox{eff}}_1(\frac{b\tau_1|\vec\nabla
   p|}{\mu_1^0})+\frac{1-\phi}{2}\mu^{\mbox{eff}}_1(\frac{b\tau_2|\vec\nabla
   p|}{\mu_2^0})
   \end{equation}
   Note that, as in Eq.~(\ref{viscosity}0, formally $\mu_{\mbox{eff}}$ is a 
function of $x$ because $\phi$ is a function of
   $x$. Darcy's law becomes
   \begin{equation} 
    \label{darcyphi} \vec u = -\frac{b^2}{12\mu_{\mbox{eff}}(\phi)}
   \left [ \vec\nabla p + \sigma\kappa(\phi)\frac{\vec\nabla\phi}{2} \right ],
    \end{equation}
    where $\kappa(\phi)$ is the curvature of the interface computed using 
   the standard expressions
     \begin{equation}
	\label{curvatureandnormal}
	\kappa(\phi)=\vec\nabla\cdot\hat r(\phi)\;\;{\rm and}\;\; \hat 
r(\phi)=\vec\nabla\phi/|\vec\nabla\phi|.
     \end{equation}
Note that $\kappa(\phi)$ is now defined in the entire space; 
$\hat r$ is the local normal to the $\phi$ isosurface. Equation
     (\ref{incompressibility}) for the incompressibility of the flow is not
     modified 
     by the introduction of the phase field. Now, the flow problem
     is completely written in terms of the phase field. 
     
To complete the model, we have to introduce an evolution equation for 
the phase field. This evolution equation should have for solution 
a smooth interface that is advected by the flow 
To this purpose, we use the equation 
presented in \cite{Folch99a} and extended with success 
in \cite{Biben03a,Biben03b} to the case of vesicles: 
      \begin{equation}
    \label{pfeqdim}
    \tau_\phi (\partial_t\phi + \vec u \cdot\vec\nabla\phi)=
    f(\phi) +w^2\nabla^2\phi-w^2\kappa(\phi)|\vec\nabla\phi|,
    \end{equation}
    with $f=\phi(1-\phi^2)$ the oposite of the derivative  of the double well
    potential $-\phi^2/2+\phi^4/4$, $\tau_\phi$ a relaxation time, and $w$ a 
small parameter that determines the width of the interface. 
In order to give a clear
   view of the equation, we first consider an oversimplified version of it with
   neither the flow nor the curvature term, in a one-dimensional space.
   The stationary solutions of this equation 
are either the uniform solutions $\phi=± 1$
or a front between a region where $\phi=1$ and a region where 
$\phi=-1$:
   \begin{equation}
   \label{tanh}
    \phi=\tanh \frac{r}{w\sqrt{2}}.
    \end{equation}
   Here, the signification of $w$ appears clearly: it is the width of
   the interface. Now let us consider this equation (still without flow and 
without the curvature term) in two dimensions. Using a perturbation method, one can 
show that
    a weakly curved interface (radius of curvature $\rho\gg w$) between
	$\phi=1$ and $\phi=-1$ is moving with a normal velocity proportional to
	$1/\rho$, the curvature of the interface. While this behaviour is
	expected in the case of phase transitions with non-conserved order
      parameters, here it is unphysical. In order 
	to suppress  this phenomenon, following \cite{Folch99a} we add the 
curvature
	term which at dominant order is the exact opposite of the term induced by
	the Laplacian when considering a curved interface. Indeed, it can be
	shown that up to the third order in $w/\rho$, the term 
      $\nabla^2\phi-\kappa |\nabla \phi|$ is equal to the unidimensional
	Laplacian computed along the axis normal to the interface. Hence,
	using $\phi=\tanh \frac{r-\rho}{w\sqrt{2}}$ (with $r$ the distance
	from the center of the interface), the right-hand side of 
Eq.~(\ref{pfeqdim}), 
      i.e. the driving force leading to unwanted
	interface movement, is equal to zero up to that third order.

       Finally, adding the term $\vec u\cdot \vec\nabla \phi$ makes the 
interface to be advected by the flow. Therefore, the dynamics of the phase field 
can be separated into two parts: a \textit{passive} part that corresponds to 
the
       advection due to the flow and an \textit{active} part that aims at
       restoring the hyperpolic tangent profile through the interface but does
       not bring any noticeable dynamics to the interface. With this principle 
in
       mind, it is clear that the relaxation time $\tau_\phi$ of the phase field 
must be fast
       enough so that the advection does not affect significantly the
       equilibrium profile. The particular choice of $\tau_\phi$ is discussed
       later.   

       Now, that model equations have been written down, we want to stress that
       while the \textit{distribution formulation} of the viscous fingering
       problem is just another way of writing down the same
sharp-interface equations,
       the phase-field model is only an approximation to them.
       To be more specific, the phase field model 
       introduces an additional length scale $w$, the interface thickness,
       which is a model parameter supposed to be small. To understand 
its
       meaning and the relationship between the phase-field approach and the
       sharp interface model, one can use the technique of matched
       asymptotics. Different  asymptotic 
       expansions of the phase field equations in powers of $w$ valid in the 
bulk
       phases and through the interface, respectively, are written down. Then, 
       matching them order by order, at dominant order in $w$ the original 
       sharp interface-problem is retrieved, which indicates that the
	results of the phase-field model converge toward the solution of the
	original problem when $w\to 0$ (the so called \textit{sharp-interface
	limit}). In other words, the model is at
	least asymptotically correct. 
	
	However, in numerical simulations, the value of $w$ should be 
      significantly larger than the space discetization and must remain finite. 
	Therefore, to be able to retrieve quantitatively correct results, one
	needs to  control the spurious effects introduced by the finite
	interface thickness and the convergence of the model  toward the
	sharp-interface limit. This can
	be done by considering the next order in $w$ in the matching
	procedure \cite{Karma98,Almgren99,Folch99a,Echebarria04}.
	Then, new $w$-dependent terms  are added to the sharp interface
	equations (this next order in the expansion is called the 
      \textit{thin-interface limit}). They actually signal the departure 
	from the $w\to 0$ limit
	and are the effect of the presence of the extra length scale.
	Physically, one expects their importance to depend on the ratio of $w$
	to the smallest genuine length scale present in the original 
      sharp-interface model. This hypothesis can then be checked by simulations
	with decreasig values of that ratio \cite{Karma98,Folch99b,Echebarria04}. 

	Here, we have written our model so that, in the case of Newtonian
	fluids, it is mathematically equivalent to the one
	presented in \cite{Folch99a}. The reason for this is that contrary to
	other phase field models \cite{Lee02,Hernandez-Machado03} for viscous
	fingering, the asymptotic expansions of this model have been
	established \cite{Folch99a} and the numerical convergence has been
	checked by considering situations where the sharp interface solution is
	well known \cite{Folch99b}. Therefore, we are confident that
	unexpected finite interface thickness effects could only arise
in our simulations in conjunction with the new feature here:
the non-Newtonian character of the more viscous fluid.

\subsection{Dimensionless equations}
\label{dimlesseqs}

In order to nondimensionalize our equations, we first look at the
relevant physical scales present in the flow, and then use the same
scales to nondimensionalize the phase-field equation. Non-dimensionalized
quantities will be denoted by a tilde. In a first
step, we define dimensionless effective viscosity functions
$\mueffitil_i$ by dividing the effective viscosity laws of the
two fluids by their zero-shear limit values $\mu_i^0$,
\begin{equation}
\label{munondimensionalization}
\mueffi_i(b\tau_i|\vec\nabla p|/\mu_i^0)=
   \mu_i^0\mueffitil_i(b\tau_i|\vec\nabla p|/\mu_i^0).
\end{equation}
Next, since in a phase-field model there is a generalized effective
viscosity valid throughout the system [Eq. (\ref{interpolatedviscosity})]
which interpolates between the effective viscosities of each fluid, we
need to choose a single viscosity scale. This choice has to be adapted
to the physical situation that is investigated. Here, we are mainly
interested in the setting used in most experiments, where the more
viscous fluid 1 is a shear-thinning liquid and the less viscous
fluid 2 is air, that is, a Newtonian fluid of very low viscosity.
Therefore, in the following we will nondimensionalize the effective
viscosity by the zero-shear viscosity of fluid 1, $\mu_1^0$. Since
fluid 2 is Newtonian, we have $\mueffi_2\equiv \mu_2^0$.

With the above choices, the nondimensionalized effective viscosity
function becomes
\begin{eqnarray}
\label{viscincontrastvariables}
      \muefftil(\phi)
=
\frac{\mueff(\phi)}
{\mu_1^0}
=
\frac{1+\phi}{2}\mueffitil_1
+\frac{1-\phi}{2}\ratio,
\end{eqnarray}
where $\ratio$ is the ratio of the two zero-shear viscosities,
\begin{equation}
\label{viscratio}
\ratio=\frac{\mu_2^0}{\mu_1^0}.
\end{equation}
This ratio can be simply related to the quantity
\begin{equation}
\label{viscositycontrast}
c\equiv\frac{{\mu}^0_1-{\mu}^0_2}{{\mu}^0_1+{\mu}^0_2}=
\frac{1-\ratio}{1+\ratio},
\end{equation}
the so-called viscosity contrast (at zero shear), also widely used in the
literature~\cite{Tryggvason83,Folch99a}.

We furthermore measure velocity in units of the outflow
velocity $U_\infty$ and lengths in units of the channel width
$W$. The natural scale for the pressure gradient that arises from the
Newtonian limit of Darcy's law is $12\mu_1^0 U_\infty /b^2$.
This yields the new dimensionless quantities
\begin{eqnarray}
\label{adimxynablakappa}
x\to \tilde{x} W \qquad y\to \tilde{y} W  \qquad
\vec\nabla\to \frac{1}{W}\tilde{\vec\nabla} \qquad
\kappa(\phi)\to \frac{\tilde\kappa(\phi)}{W} \\
\label{adimumup}
{\vec u} \to {\tilde{\vec u}} U_\infty \qquad
 \vec\nabla p \to \frac{12\mu_1^0 U_\infty}{b^2}\tilde{\vec\nabla}\tilde{p} \\
\label{adimt}
t \to \tilde{t} \frac{W}{U_\infty}.
\end{eqnarray}
Under this change of variables, the arguments of the dimensionless
effective viscosity function given by 
Eqs.~(\ref{munondimensionalization},\ref{viscincontrastvariables})
become
\begin{eqnarray}
\label{fullydimlessvisc}
\muefftil(\phi)
&=&\frac{1+\phi}{2}\mueffitil_1(\We |\tilde{\vec\nabla} \tilde p|)
           +\frac{1-\phi}{2}\ratio\;,
\end{eqnarray}
where the Weissenberg number $\We$ is defined by
\begin{equation}
\label{globalwe}
\We=\frac{12\tau_1 U_\infty}{b}.
\end{equation}
In the remainder of this paper (except for Appendix~\ref{sec_darcy}), we
will work in these new dimensionless variables and drop the tildes
for simplicity.

The incompressibility condition remains formally the same,
and the dimensionless version of Darcy's law reads
\begin{equation}
\label{darcyless}
\vec u = -\frac{1}{\mueff(\phi,\We|\vec\nabla p|)}
\left[\vec\nabla p + \Gamma\kappa(\phi)\frac{\vec\nabla\phi}{2}\right],
\end{equation}
where
\begin{equation}
\label{Gamma}
\Gamma=\frac{b^2\sigma}{12W^2 \mu_1^0 U_\infty}
\end{equation}
is a dimensionless surface tension. In summary, the flow equations
contain three dimensionless parameters: 
the dimensionless surface tension $\B$,
the Weissenberg number $\We$,  and the viscosity ratio $\ratio$. 
A more detailed discussion of these parameters
and their role in the finger selection process is deferred to
Sec.~\ref{sec_params} below.

To complete the set of dimensionless equations, we apply the same
scaling to Eq.~(\ref{pfeqdim}) for the phase field. We obtain
\begin{equation}
\label{pfeqless}
\frac{\tau_\phi U_\infty}{W}(\partial_t\phi +
\vec u \cdot\vec\nabla\phi)=
f(\phi) +(\frac{w}{W})^2[\nabla^2\phi-\kappa(\phi)|\vec\nabla\phi|],
\end{equation}
and identify the dimensionless interface thickness $\epsilon=w/W$,
the ratio of the interface thickness to the channel width.
In order to reduce the number of purely computational parameters,
we choose $\tau_\phi=\epsilon w/U_\infty$. Indeed, $w/U_\infty$ is the
time it takes a flow of the magnitude of the base flow $U_\infty\hat y$
to cover one interface thickness $w$, and the extra small $\epsilon$
factor ensures that the phase field relaxation is one order in
$\epsilon$ faster than the forcing by the flow. We finally get
\begin{equation}
\label{pfeq}
\epsilon^2\partial_t\phi =
f(\phi) - \epsilon^2\left[\nabla^2\phi-\kappa(\phi)|\vec\nabla\phi| -
\vec u \cdot\vec\nabla\phi\right].
\end{equation}

\subsection{Incompressibility and boundary conditions}
\label{incompress}

In the simulations, Eq.~(\ref{pfeq}) for the phase field
and the fluid flow equations need to be solved simultaneously.
The fluid flow part, in turn, implies solving Eq.~(\ref{darcyless})
taking into account the incompressibility condition,
Eq.~(\ref{incompressibility}). There are several ways to
implement incompressibility.

One possibility is to take the curl of Eq.~(\ref{darcyless}),
which eliminates the pressure field in the Newtonian case.
Incompressibility is equivalent to the requirement that the
flow is potential, that is, the velocity field can be written
as derivatives of the stream function. The curl of Eq.~(\ref{darcyless})
yields a Poisson equation for the stream function. This strategy leads 
exactly to the model of
Ref.~\cite{Folch99a} for Newtonian fluids, as desired. 

However, for non-Newtonian rheologies, the dependence of the
effective viscosity on $|\vec\nabla p|$ implies that the pressure
cannot be eliminated in this straightforward manner any more.
Therefore, we use here a velocity-pressure formulation:
We take the divergence of Eq.~(\ref{darcyless}) and use the
incompressibility condition, which yields
\begin{equation}
\label{pressure}
\vec\nabla\cdot\left
[\frac{\vec\nabla p}{\muefftil(\phi,\We|\vec\nabla p|)}\right ] =
- \vec\nabla\cdot\left [
\frac{\B\kappa(\phi)\vec\nabla\phi}{2\muefftil(\phi,\We|\vec\nabla p|)}\right ].
\end{equation}
For a given configuration of the phase field $\phi$, Eq.~(\ref{pressure}) 
together with appropriate boundary conditions (discussed below) completely
specifies the pressure field $p$. In the Newtoninan limit where
the effective viscosity is pressure-independent, this equation is
a Poisson equation for the pressure inside the interfacial regions
where the phase field $\phi$ varies, and reduces to the Laplace
equation in each bulk domain. In the non-Newtonian case, the
source term is present also in the bulk, and an iterative Poisson
solver must be used to obtain the pressure field
for the given configuration of the phase field at each timestep.
Then, the original Eq.~(\ref{darcyless}) immediately yields
the velocity field $\vec u$. This is then used in the next
time step to advect the phase-field $\phi$, as prescribed 
by Eq.~(\ref{pfeq}). More details about the numerical procedure
are given in Appendix~\ref{sec_numerics}.

Furthermore, boundary conditions for the phase and pressure fields 
are required at the edges of the channel. For simplicity, we will 
assume that if an interface crosses any of the boundaries, it will 
do so at a 90$^\circ$ angle, which implies that the derivatives of 
the phase-field normal to the boundaries are zero
(reflecting boundary conditions): 
\begin{eqnarray}
\partial_x \phi &=& 0\,\, (y=± L/2),\\
\partial_y\phi  &=&  0\,\, (x=± W/2)
\end{eqnarray}
 Since the lateral walls
are sealed and hence $u_x=0$, we also have 
\begin{equation}
\partial_x p=0
\mbox{ at }x=± W/2.
\end{equation}
 The only non-trivial boundary conditions are
the pressure boundary conditions at the inlet and
the outlet, where either the pressure or its gradient have
to be prescribed. Since we have considered a flow with a
fixed overall flow rate, we should prescribe 
the pressure gradient.

At the outlet, only fluid 1 is present.
If the interface remains far enough from the outlet, the pressure
is simply a constant along the entire outlet, and the pressure
gradient is directed along the $y$ direction.
Since, at the outlet, the dimensionless velocity is equal to
$(0,1)$ (corresponding to a uniform flow with velocity $U_\infty$
along the $y$ direction), the Darcy law (eq. \ref{darcyless}) implies that  the 
pressure
gradient is the solution of the equation
\begin{equation}
\label{eq_pboundaryoutlet}
|\partial_y p| = \mueff (\phi=+1,\We|\partial_y p|).
\end{equation}
where the velocity $U_\infty$ enters the equation through the
Weissenberg number. This equation can be solved numerically in 
a straightforward way. In our simulations, we 
start with an initial guess for the pressure gradient, which is
then updated at  each time step with the value found by the pressure 
solver in the vicinity of the outlet. This procedure rapidly converges to
the fixed point which is the solution of Eq.~(\ref{eq_pboundaryoutlet}).

As for the inlet, we  consider the  case where both fluids are present. For
a well-developed steady-state Saffman-Taylor finger, the sides of
the finger are parallel to the channel walls up to a correction that
decays exponentially with the distance from the finger tip. Therefore,
if a sufficiently long portion of the finger is inside the simulation
box, the interfaces that cross the inlet can be considered flat and
normal to the boundary, and the fluid velocity along the
$x$ direction is zero in both fluids. Therefore, there
is no pressure gradient along the $x$ direction, which of 
course implies that the pressure gradient is directed 
along $y$, and constant along the inlet.

In contrast, the fluid velocity varies along the inlet,
since the viscosity does change when crossing the interface.
However, its integral along $x$, $\int_{-1/2}^{+1/2} u_y dx$, which represents
the net inward flow, must be equal to unity, since the flow is
incompressible and the fluid exits the outlet at a rate of unity
in our dimensionless variables. Integrating the $y$ component of
Eq.~(\ref{darcyless}) along the inlet, we thus obtain
\begin{equation}
\label{inletpressure}
|\partial_y p| =
\frac{1}{\int_{-1/2}^{+1/2}\frac{1}
{\mueff(\phi,\We |\partial_y p|)}dx},
\end{equation}
which constitutes a closed equation for the desired value of
$|\partial_y p|$ at the inlet.

\subsection{Simulation procedure}
\label{procedure}

In our numerical studies, our main focus is on steady-state fingers.
Although we could start each simulation with a weakly 
perturbed flat interface and let it follow its natural dynamics
until a steady finger stabilizes, this is not the most 
efficient procedure for parametric studies of the finger
width as a function of $\B$ and $\We$. 
Therefore, we instead first
calculated an initial finger profile for values of the control
parameters where convergence can be easily achieved, and then
use the resulting steady-state pressure and phase fields as
initial condition for a run with slightly different control
parameters. Increasing or decreasing $\B$ and/or $\We$ in
small steps, we are thus able to follow the steady-state
solution branches over a substantial parameter range.

When performing the first computation for a given viscosity law, 
we set the initial interface profile to a semi-elliptic bubble 
(of width $W/2$ and length $W$) growing 
from the inlet of the channel. The initial configuration
of the phase field is a hyperbolic tangent profile in the 
elliptic coordinates, and its zero contour is located at the elliptic 
bubble interface. The simulations are performed in a channel
with a length of $L=5W$. The bubble increases in size and depelops
into an elongated finger. When it reaches a reference position
(typically, located at twice the channel width from the inlet), 
the whole domain is translated backward by one grid 
spacing (in other words, the finger is pulled back by 
one grid point). The velocity of the finger is computed by measuring
the time between two successive pullbacks. The finger width is 
measured at the entrance of the channel when a pullback occurs. 
We consider the steady reached when both tip velocity and finger 
width vary less than a fixed value (here chosen to be $10^{-8}$,
to be compared with a typical tip velocity of 2 and a typical 
finger width of 0.5) between two pullbacks. 

Values of $\B$ of the order of $10^{-2}$ yield 
a rapid convergence to a steady-state finger,
both for Newtonian and non-Newtonian fluids. For the latter, 
the convergence is more difficult to obtain because of the 
nonlinearities in the viscosity laws. Typically, we calculate
the first finger with a low value of the Weissenberg number $\We$
for which these nonlinearities are small; $\We$ is then increased
progressively up to the desired value. In this way, values up to
$10^2$ can be treated, for which the pressure solver would have
otherwise not converged. As for $\B$, the values we are able to attain
are limited both from below and from above. For small values of
$\B$, results become sensitive to the discretization and the
interface thickness, as will be detailed below. For large
values of $\B$, the finger width becomes close to unity, and
the tail of the phase-field profile starts to interact with
the sidewalls.

The solutions found in our simulations are single fingers 
propagating at constant velocity along the channel. 
We consider fingers symmetric with respect to the channel mid-line
$x=0$. This allows us to reduce the 
computation time by limiting the numerical domain to half 
the channel: \(0<x<1/2\)). 
The validity of this procedure was checked by occasinally 
performing computations in the full domain: fingers started 
with an axis of symmetry shifted away from the mid-line always 
relax towards the center of the 
channel in finite time. We have also checked that increasing
the length $L$ of our simulation domain (changing the aspect ratio
$L/W$) does not change the results.
Indeed, in our typical steady-state configuration, the back
of the finger is cut off at twice the channel width behind
the tip, where its flanks are almost flat and fluid 1 is almost at
rest. Furthermore,
the pressure field becomes almost linear far ahead of the tip, 
and a distance of three times the channel width is enough
to resolve all non-trivial features of the velocity and 
pressure fields.

In our simulations we let the 
finger extend inside the channel until the tip crosses a reference position 
along the \(y\) axis. When this happens, the time step is truncated so that 
the finger tip advances exactly to the pullback coordinate; the whole field is 
then pulled one grid step backward. The velocity of the finger is obtained by 
computing the average velocity between two successive pullbacks. The finger 
width is measured at the entrance of the channel when a pullback occurs. The 
stationary state is declared to be achieved when both tip velocity and finger 
width vary less than a fixed value, here chosen to be \(10^{-8}\).

The computation time necessary to achieve the stationary state for a given 
\(B\) value can be significantly reduced when the run is initialized with a 
finger profile close enough to the converged state. Hence we applied the 
following procedure to obtain selection curves in the Newtonian and the shear 
thinning cases:

For the first computation of the set, the phase field is initialized with a 
semi-elliptic bubble growing from the inlet side of the channel. The small 
radius spans over half the width of the channel and the big radius is 
arbitrarily chosen to be the channel width in the longitudinal direction. 
The phase field obeys a hypertangent profile in elliptic coordinates. A moderate 
\(B\) parameter of \(10^{-2}\) is chosen to compute the first finger profile.
The parameter \(B\) is then varied towards zero and towards infinity to 
move along the selection curve. Each computation is initialized with the 
finger profile at numerical convergence.
Attainable \(B\) values are both limited in the small and large limits. In the 
former case, the interface thickness needs to be reduced, and thus the grid 
refined, in order to retain the relevant selection mechanism. In the latter 
case, the stationary solution can be destroyed when the phase field is too close 
to the boundaries.
In the non-Newtonian case the first step is more difficult because of the 
nonlinearities in the viscosity. We do not impose the correct pressure condition 
at the outlet, but rather let it relax as time is stepped forward.
\subsection{Control parameters and finger selection}
\label{sec_params}

The independent parameters that appear in our equations are
the zero-shear viscosity ratio $\ratio$
(constant for a given pair of fluids),
the Weissenberg number $\We$, which controls the intensity
of the shear-thinning effect,
and the dimensionless surface tension $\B$. It is noteworthy
that in experiments performed with a single Hele-Shaw cell of
fixed width and gap spacing,
both $\We$ and $1/\B$ increase linearly with $U_\infty$ [see
Eqs. (\ref{globalwe}),(\ref{Gamma})],
which is the only parameter that can be externally controlled.
The full two-dimensional parameter space can hence only be
explored in experiments by varying the channel geometry as 
well as $U_\infty$.
In contrast, in the simulations it is easy to vary these two
parameters independently, and to determine the selected finger
width. However, it is useful to take some additional 
considerations into account.

It is known that two main ingredients determine the finger 
width: the dimensionless surface tension and the anisotropy of
the interface or the medium. In shear-thinning fluids an 
effective anisotropy arises from the fact that the in-plane
velocity and thus the shear are maximal at the tip, and decay
when going to the sides of the finger. As a consequence, the viscosity
and hence the mobility in Darcy's law vary along the interface.
Thus, the strength and nature of this effective anisotropy 
are essentially controlled by the Weissenberg number and
the functional form of the viscosity law.

Let us now turn to the dimensionless surface tension.
For Newtonain fluids, it was shown \cite{McLean81}
that the selection of the finger width is determined by a single
dimensionless parameter $B$, which represents 
the ratio of stabilizing (capillary) to destabilizing (viscous)
forces. The latter are proportional to the finger speed and the
{\em difference} of the two viscosities 
(see e.g. Ref. \cite{Tryggvason83}). 
In shear-thinning fluids, 
the relevant viscosity is the one in the vicinity of the tip,
and the correct definition of the parameter $B$ is
\begin{equation}
B = \frac{b^2\sigma}{12W^2 U_{\rm tip}\,
   [\mu_1^0\mueffi_1(\We|\vec\nabla p_{\rm tip}|)-\mu_2^0]}.
\label{eq_btipdef}
\end{equation}
Using the definition of $\B$ and the fact that 
mass conservation for an incompressibe fluid enforces $U_\infty=\lambda U_{\rm 
tip}$
for a steady-state finger of relative width $\lambda$, we find the following relation 
between $B$ and $\B$:
\begin{equation}
B = \frac{U_\infty}{U_{\rm tip}}
          \frac{\B}{\mueffi_1(\We|\vec\nabla p_{\rm tip}|)-\ratio}
  = \frac{\B\lambda}{\mueffi_1(\We|\vec\nabla p_{\rm tip}|)-\ratio}.
\label{eq_Bgammarel}
\end{equation}

Ideally, we would like to explore the parameter space along lines 
of constant $B$ in order to track only the influence of the effective 
anisotropy (the selection parameter of the isotropic Saffman-Taylor 
problem is then constant). 
However, $B$ is difficult to control directly in 
our simulations: the effective viscosity at the tip, which is
needed to calculate $B$, depends on the finger speed, which
is itself the result of the selection to be investigated.
Therefore, we explore the width selection by varying either
$\We$ at fixed $\B$, or $\B$ at fixed $\We$, and calulate 
$B$ {\em a posteriori} using the tip speed
$U_{\rm tip}$ and pressure gradient $|\vec\nabla p_{\rm tip}|$
extracted from the simulations. Note that this procedure
is perfectly analogous to the one followed in experiments:
the viscosity at the tip is estimated {\em a posteriori}
using the measured finger speed \cite{Lindner00,Lindner02}.
Keeping $B$ constant is more involved, and would require some 
iterative trial and error procedure, which is perfectly 
feasible but cumbersome.

A last point that deserves brief mention is the viscosity
ratio $\ratio$. In the case of air pushing a viscous fluid, $\nu$ is
extremely small, so that the viscosity of the air can be
neglected altogether. In our numerical formulation, however,
it is difficult to simulate very small values of $\ratio$,
because Eq.~(\ref{pressure}) then has extremely different numerical
stiffness in the two bulk domains, which makes the convergence
of the pressure solver delicate. In our simulations, we have
typically used values of $\ratio$ ranging from $5× 10^{-2}$
to $5× 10^{-4}$, which are large enough to guarantee a
robust and efficient solution of Eq.~(\ref{pressure}).
One could think that these are small enough to neglect 
$\ratio$ in the denominator of Eq.~(\ref{eq_Bgammarel}). 
However, as will be seen below, for a viscosity law 
without lower bound (such as a power-law),
the viscosity of the shear-thinning fluid
will become comparable to or even smaller than that of the 
pushing Newtonian fluid even for $\ratio \ll 1$, for sufficiently 
high Weissenberg numbers. In the latter case, the fingering
instability disappears altogether. We insist that this is an 
entirely physical effect that should be experimentally 
observable in fluid couples of not too different viscosities.

\section{Results}
\label{results}

\subsection{Newtonian fluid}
\begin{figure}[htbp]
\centering
\includegraphics[width=\textwidth]{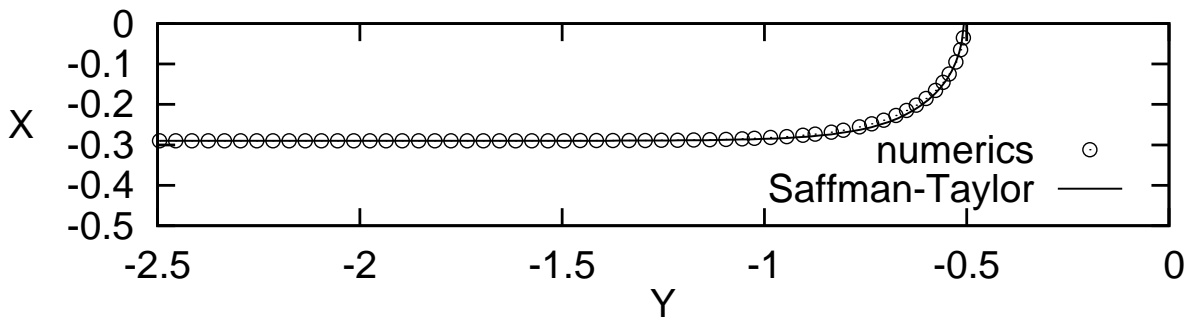}
\caption{Comparison of computed interface
( \(\B=0.01\), \(\ratio\)=0.05,
\(\epsilon\)=0.02, $\Delta x$=0.01) 
 and analytical solution of
Saffman and Taylor, \(\lambda=0.58\).}
\label{fig_prof_num_an}
\end{figure}

\begin{figure}[htbp]
\centering
\includegraphics[width=.8\textwidth]{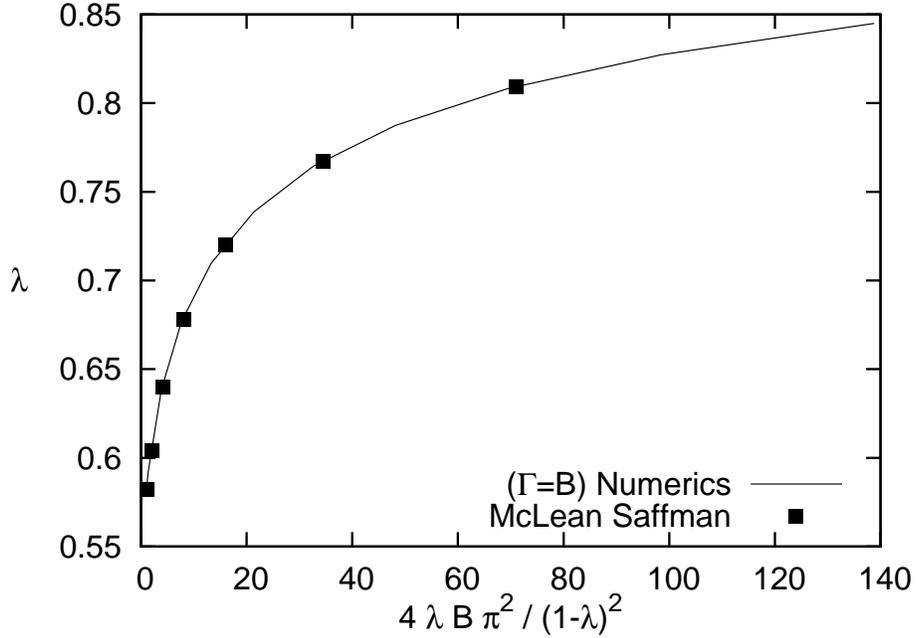}
\caption{Comparison of computed finger width $\lambda$ 
(\(\ratio=5×10^ {-3}\),
\(\epsilon\)=0.02, $\Delta x$=0.01)
and semi-analytical solution of McLean and Saffman.}
\label{fig_Wnewt_140_700_co9_mclsf}
\end{figure}

In order to test our model formulation and its numerical implementation,
we start by performing simulations in Newtonian fluids. The simulations 
converge without difficulty to a steady-state finger solution.
In Fig.~\ref{fig_prof_num_an} we display a comparison between
a typical finger shape extracted from our simulations and the
analytical solution of Saffman and Taylor \cite{Saffman58},
\begin{equation}
x=\frac{\lambda}{2\pi}
\arccos\left[2\exp\left(2\pi \frac{y-y_{tip}}{1-\lambda} \right)-1\right].
\label{eq:STshape}
\end{equation}
After fitting $\lambda$, the agreement between the computed
and analytical curve is good. There are some small discrepancies
close to the finger tip that are to be expected, since the solution
given by Eq.~(\ref{eq:STshape}) does not contain the effect of
surface tension. That the latter is correctly incorporated into
our model is proven by the results shown in
Fig.~\ref{fig_Wnewt_140_700_co9_mclsf}, where we display the
selection curve for the finger width at fixed
values of $\ratio$ and $\epsilon$ as a function of the dimensionless
combination of parameters $4\lambda B\pi^2/(1-\lambda)^2$ used in the classical
work of Mc Lean and Saffman \cite{McLean81}, which we compare to.
The agreement is excellent, except for very small values of $\B$.
This constitutes an extremely sensitive test for our model since
the finger width is selected by the surface tension ({\em via} the
selection parameter $B$) through a singular perturbation mechanism.

\begin{figure}[htbp]
\centering
\includegraphics[width=.8\textwidth]{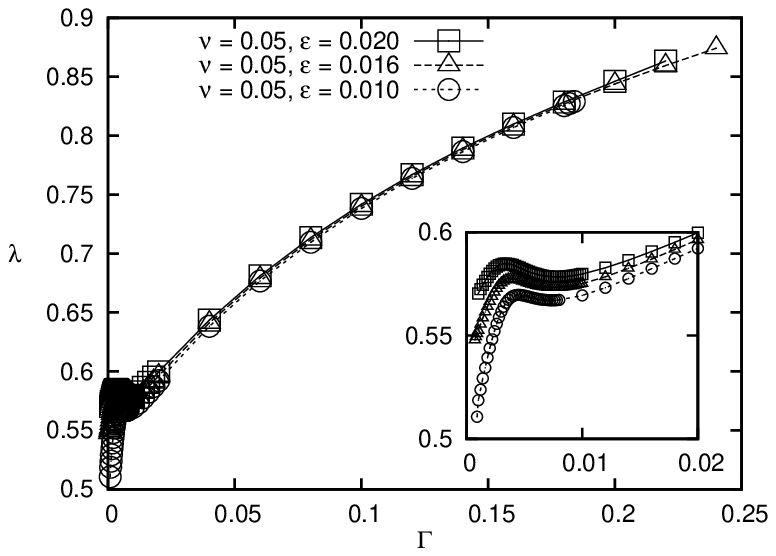}
\includegraphics[width=.8\textwidth]{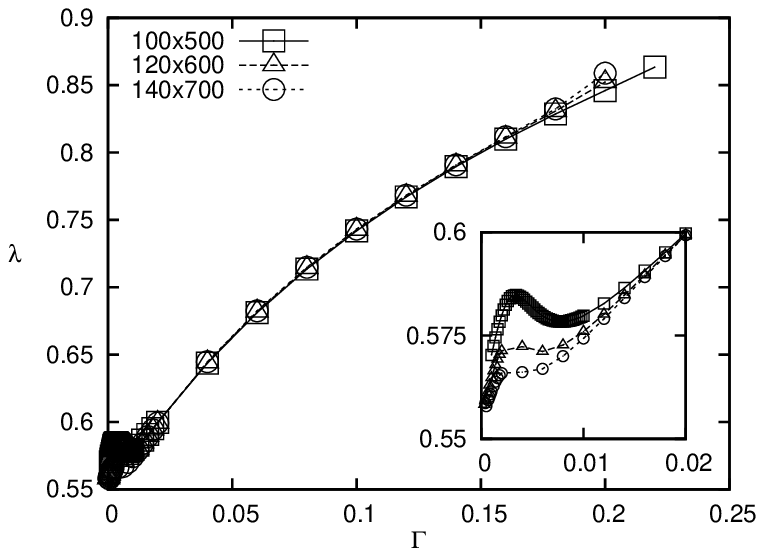}
\caption{Steady-state finger width versus \(\B\), (a) for various
\(\epsilon\)=0.02, 0.016, 0.01 and $\Delta x$=0.01, 0.008, 0.005, 
and (b) for fixed $\epsilon$=0.02 and
increasing resolution of discretization, $\Delta x$=0.01, 0.008333,
0.005. \(\nu\)=0.05.}\label{fig_Wnewt_co9_comp}
\end{figure}

In Fig.~\ref{fig_Wnewt_co9_comp} we replot the selection curve
directly as a function of $\B$, 
to make the deviations from the analytical prediction 
for small values of $\B$ most apparent.
These deviations take place for $\B<0.01$;
Below that value, the decrease of the finger width $\lambda$ with
the dimensionless surface tension $\B$ to the predicted limit value
of $\lambda=0.5$ for $\B\to 0$ (note that for a Newtonian
fluid, $B$ is just proportional to $\Gamma$) is interrupted by a small ``bump''. 
Two effects limit the precision of our results.
First, it is expected that at
low values of $B$ smaller values of $\epsilon$ are needed to
obtain properly resolved results. The reason is that the
wavelength of the marginally stable mode of the linear
Saffman-Taylor instability scales as $\sim \sqrt B$. As in
any phase-field model, the correct interface dynamics can
only be guaranteed {\it a priori} when $\epsilon$ remains smaller than this
value (i.e., well into the thin-interface limit). 
Deviations from the sharp-interface solution are thus
simply a sign of insufficient resolution of the relevant
length scale by the phase field. The second effect is purely 
numerical: when $\B$ is decreased the surface tension 
effect becomes numerically small. More precisely, the pressure 
gradient accross the interface created by the Laplace pressure 
becomes smaller and smaller with respect to the global driving
pressure gradient. Therefore, discretization errors can become 
significant. In particular, the anisotropy induced by the 
discretization on a regular lattice can have a strong effect 
on the solution. This is especially critical, since it is
known that even a small amount of interfacial anisotropy
dramatically modifies the selection mechanism \cite{Kessler86,Dorsey87}.

In Fig.~\ref{fig_Wnewt_co9_comp}, we test the importance of
these two effects. Whereas a reduction in $\epsilon$ at fixed
resolution (that is, constant $\epsilon/\Delta x$) reduces
the height of the ``bump'', the change of sign in slope occurs always at
similar values of $\B$. In contrast, if the mesh is refined at fixed 
$\epsilon$, the change in slope is shifted towards smaller values
of $\B$. This indicates that the
numerical discretization error is the dominant effect. Since
a further decrease in the grid spacing would require a much larger
computation time, we have  limited our study to the regime of intermediate
values of $\B$. 

Incidentally, an observation we find worth reporting is that of
symmetrical pulsating fingers ({\it i.e.} time-periodic 
solutions with oscillating width and tip velocity), albeit in a
parameter region where the numerical convergence is not 
guaranteed ($\B$ slightly below $(10^{-4}$).
These oscillations disappeared after further grid refinement.
This is consistent with the picture \cite{Bensimon86} according to
which the threshold in the logarithm of the amplitude of the noise
(here numerical and related to the grid)
needed to nonlinearly destabilise a Saffman-Taylor finger decays
linearly with $-\B^{-\beta}$, $\beta>0$, $\beta\sim0.5$.

\subsection{Shear-thinning fluids}
To study the effect of shear thinning, we first need to specify the viscosity
law. As an example, we take a two-plateau Carreau fluid, whose viscosity obeys the
equation
\begin{equation}
\label{carreau2plateaux}
\frac{\mu(\tau\dot\gamma)-\mu^{\infty}}{\mu^0-\mu^{\infty}}=(1+(\tau\dot\gamma)^2)^{(n-1)/2}.
\end{equation}
Besides the already introduced relaxation time $\tau$ 
and zero-shear viscosity $\mu_1^0$,
this law has an inifinite-shear asymptote at the value $\mu^{\infty}$
and an exponent $n$.
It describes three regimes: two Newtonian plateaux at zero and
infinite shear, where the viscosity is independent
of the shear rate, and a shear-thinning region in between.
The ratio of the heights of the two plateaux can be defined,
$\alpha=\mu_1^\infty/\mu_1^0$, whereas the slope in the shear-thinning regime is
determined by both $n$ and $\alpha$.

In the following, we address two limiting cases of this general law:
the weakly ($\alpha$ not too small, see below for a more precise statement)
and the strongly ($\alpha\to 0$) shear-thinning regimes.
No analytic expression for the corresponding
effective viscosity (to be used in Darcy's law) is known
in either limit.
\subsubsection{Weakly shear-thinning fluids}

We first consider the weakly shear-thinning case 
and set $n=-1$ in Eq. (\ref{carreau2plateaux})
to make contact with Ref.~\cite{Fast01}.
It was shown there that the resulting law translates into an
{\em effective} viscosity in Darcy's law as long as $\alpha>1/9$, so 
for practical purposes that sets the minimal value of $\alpha$ which we mean
when we refer to the ``weakly'' shear-thinning case.
However, no closed analytical expression for this effective viscosity seemed
possible, but the same functional dependence as
the viscosity law Eq. (\ref{carreau2plateaux}) with $n=-1$
turned out \cite{Fast01} to provide an excellent approximation for it:
\begin{equation}
\label{2plateaux}
\mueffi_1=\frac{1+\alpha\,|\We\vec\nabla p|^2}{1+|\We\vec\nabla p|^2},
\end{equation}
where we recall that the Weissenberg number is given by 
$\We= 12 \tau_1U_\infty/b$. 
We therefore use this law in the
remainder of this section.

The value $\alpha=1$ corresponds 
to a Newtonian fluid; when \(\alpha\) decreases, the viscosity 
variations become steeper. 
We recall that there are now two independent 
parameters that control the finger selection (on top of $\alpha$):
$\B$, as in Newtonian fluids, 
and $\We$, which measures the strength of the shear-thinning effect.
We begin by investigating the role of $\We$.

\begin{figure}[htbp]
\begin{center}
\includegraphics[width=.8\textwidth]{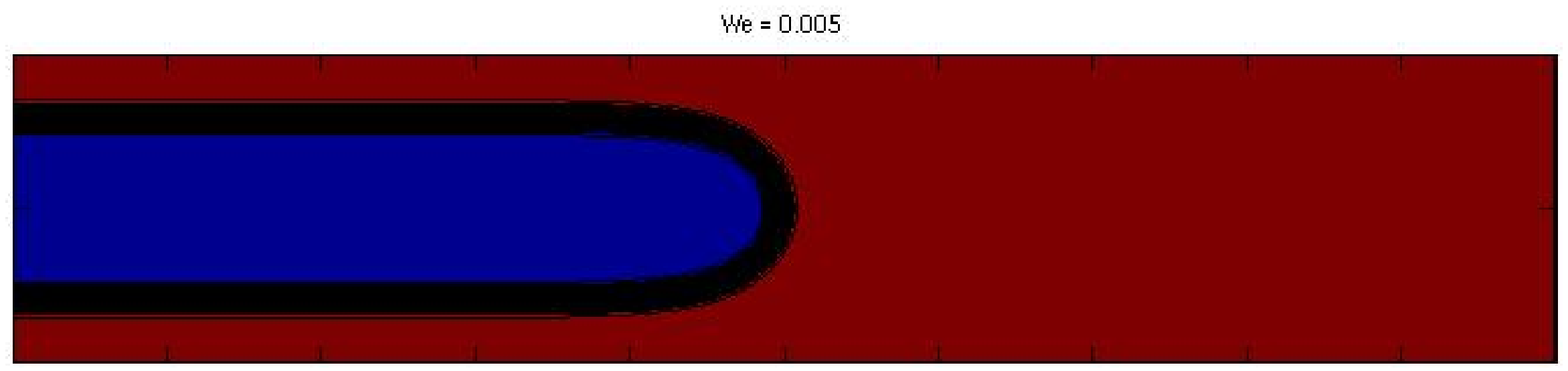}
\includegraphics[width=.8\textwidth]{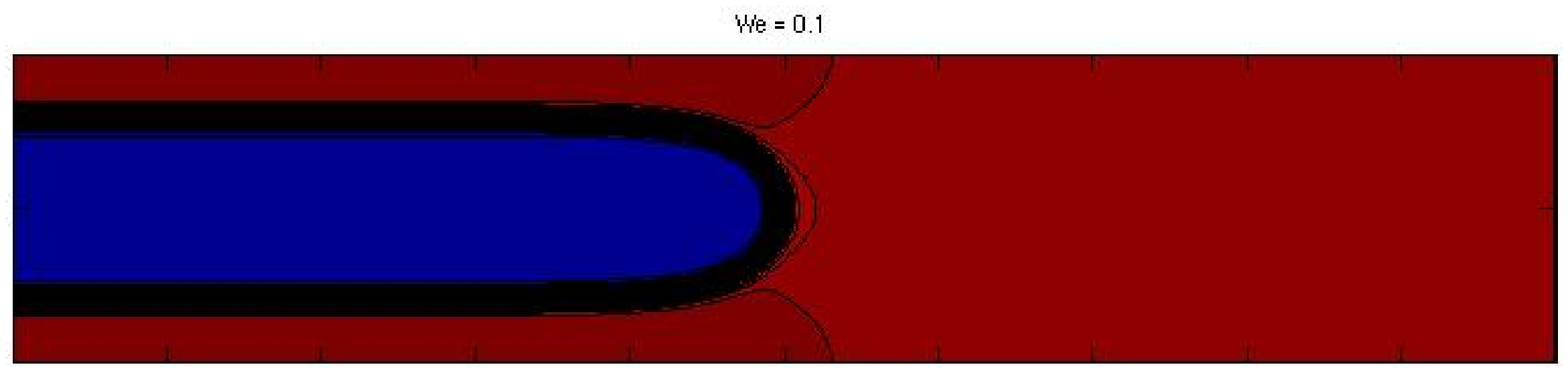}
\includegraphics[width=.8\textwidth]{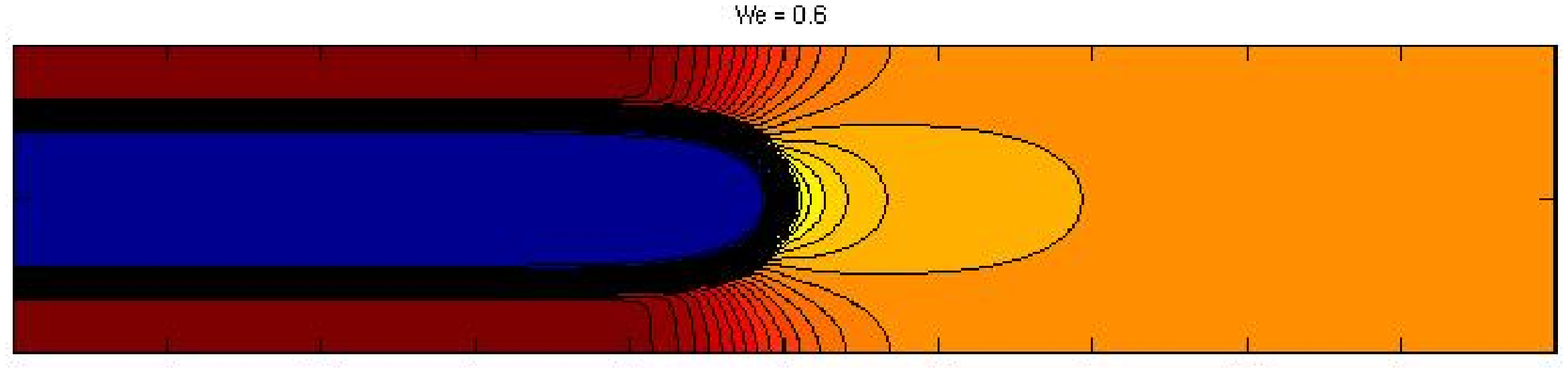}
\includegraphics[width=.8\textwidth]{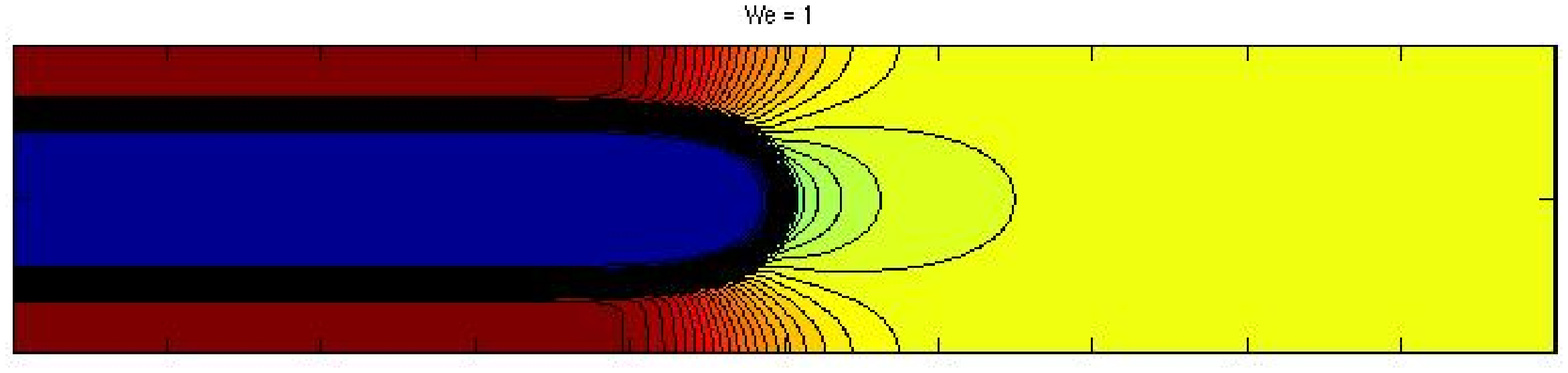}
\includegraphics[width=.8\textwidth]{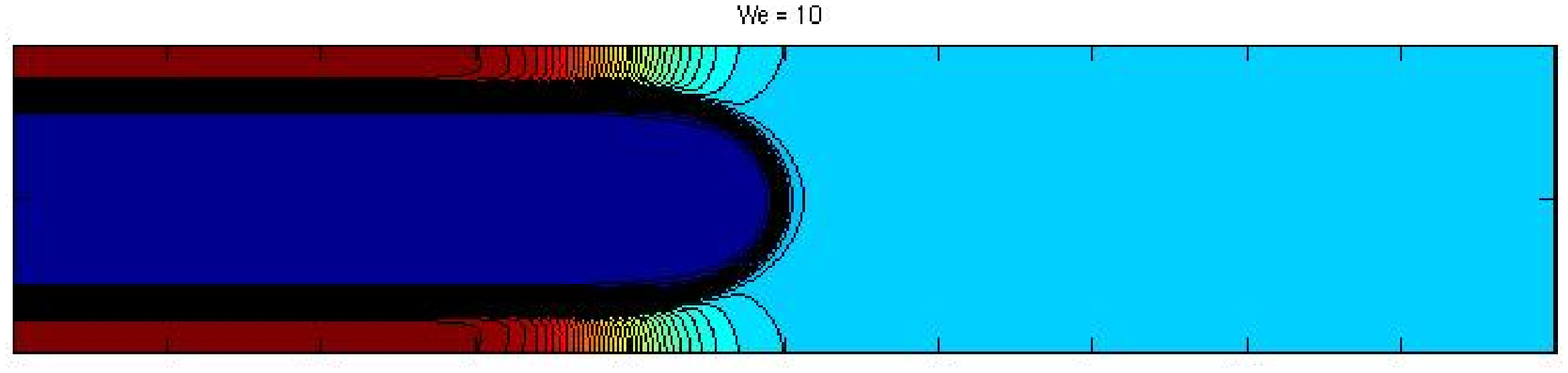}
\includegraphics[width=.8\textwidth]{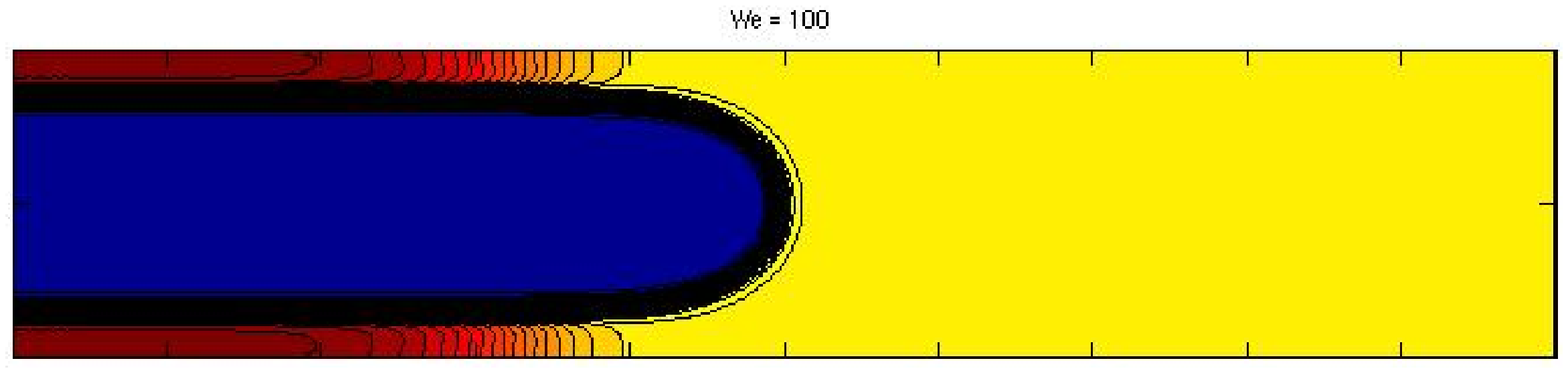}
\end{center}
\caption{Maps of the viscosity for the two-plateau law
with $\alpha=0.3$ at various Weissenberg numbers. 
Darker tones correspond to more viscous regions.}
\label{fig:flowregimes}
\end{figure}

Let us first illustrate the origin of the effective anisotropy 
effect for shear-thinning fluids by display maps of the local effective 
viscosity function $\muefftil$, Fig.~\ref{fig:flowregimes}, in various 
flow regimes, {\it i.e.}, for various ranges of $\We$ values.
We find it clearer to begin with a description of the velocity field, since it
relates directly to the local viscosity through the shear
rate, which is proportional to the gap-averaged velocity. 
Far ahead of the finger, the local velocity is $U_\infty=1$ as in the
outlet. The speed increases when the finger is approached, since the
finger tip speed,
$U_{tip}=U_\infty/\lambda\sim 2$ is larger. 
Indeed, this is the maximal speed in the system. 
Further upstream (along the finger flanks) the speed
of fluid 1 decreases to its limiting value, which can be computed using
Eq.~\ref{inletpressure} and is of the order of $\nu/\lambda$. 
For $\nu=0$ (inviscid pushing fluid) the limiting value is 0. 

With this picture in mind, the shear-thinning phenomenon upon
increasing $\We$ should be clearer. 
For $\We\ll 1$, we remain in the low-shear Newtonian plateau of the viscosity,
which is hence homogeneous.
As ($\We>0.1$), the speed at the finger tip 
enters the shear-thinning regime, so the
effective viscosity exhibits a well-marked minimum there. Furthermore,
it increases towards its Newtonian limit along the finger sides,
and it also increases ahead of the finger and towards the outlet.
This picture remains valid when $\We$ increases further, 
with the only difference that the region where the Newtonian regime is reached is 
sent further upstream along the finger flanks. 
Eventually, for ($We>5$), a third 
regime appears: The fluid at the finger tip enters the high-shear-rate plateau of 
the viscosity law, so the viscosity becomes homogeneous in a growing region close to the
tip, although it remains its absoute minimum in space. 
Soon the outlet is taken by this homogeneous-viscosity region, since
the speed there is typically just a factor 2 smaller than at the tip.
However this is not the case of the finger flanks, where the fluid
speed decreases much more upstream, so they remain a shear-thinning
zone, provided $\nu$ is small enough. This shear-thinning zone expands
and moves upstream as $\We$ is furhter increased;
if $\nu>0$ is kept constant, it will eventually reach the inlet,
and if $\We$ is even increased further, the whole shear-thinning zone
will ``pass'' through the inlet until the spot reaches the high-shear plateau
and the viscosity becomes homogeneous again everywhere (but now lower).
This happens at $\We\approx 1000$ for $\nu=5.10^{3}$, regardless of the finger
length simulated.

\begin{figure}[htbp]
\centering
\includegraphics[width=10cm]{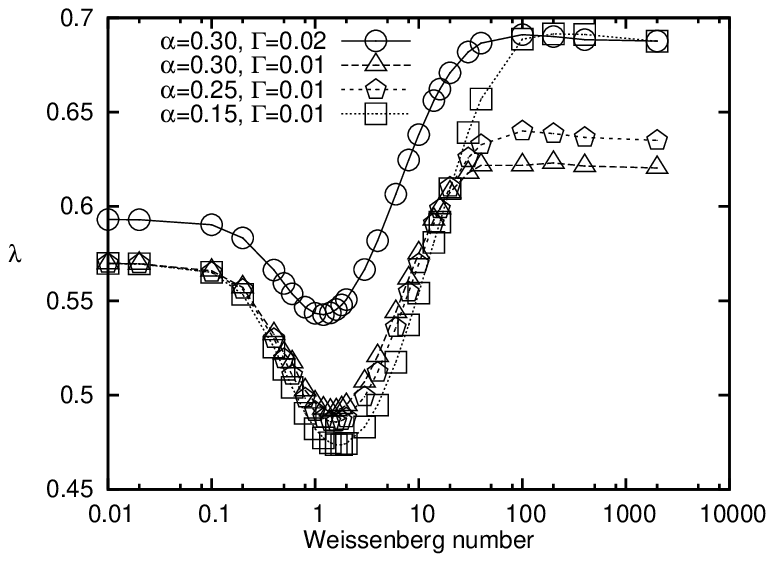}
\caption{Finger width $\lambda$ versus $\We$ at fixed values
of $\B$ and $\alpha$ for the two-plateau viscosity law.}\label{fig_width_We_a}
\end{figure}

In Fig.~\ref{fig_width_We_a}, we show the selected finger width
as a function of $\We$ at fixed $\B$ for various values of $\alpha$ 
and $\B$. For small values of $\We$, the shear-thinning fluid is 
in the high viscosity plateau
and the finger width is almost constant. When $\We$ is increased
and approaches unity, the finger width decreases. The curve goes
through a minimum, after which the finger width increases with \(\We\)
until it becomes constant again when the shear-thinning fluid
enters the second plateau.

The finger width for given $\alpha$ and $\B$ is larger
at  $\We>>1$ than at $\We<<1$. This is a consequence of the
relation between the control parameter $\B$ and the tip selection
parameter $B$ already discussed in Sec.~\ref{sec_params}:
the control parameter $\B$ is defined with the viscosity
of the first Newtonian plateau. However, at high shear rates,
the fluid around the tip is in the second plateau, and therefore
the width selection is governed by the corresponding value of
the viscosity. Neglecting the viscosity of the Newtonian fluid
(that is, setting $\ratio=0$), we obtain at
high Weissenberg numbers the simple relation $B=\B/\alpha$,
whereas for low $\We$, $B=\B$. Since $\alpha<1$, larger fingers
are selected for high $\We$. This argument is corroborated by
the two curves for $\B=0.02$, $\alpha=0.3$ and
$\B=0.01$, $\alpha=0.15$, which tend to the same finger
width at high $\We$ (Fig. \ref{fig_width_We_a}). Indeed, they have
the same value of $B=0.033$ in that regime.

A noteworthy feature of Fig.~\ref{fig_width_We_a} is that all
the curves for $\B=0.01$ exhibit finger widths that are lower
than $0.5$, which is the smallest value that can be achieved
in Newtonian fluids. This narrowing is due to the effective
anisotropy induced by the shear-thinning effect in the medium,
as can be appreciated from the viscosity maps in
Fig.~\ref{fig:flowregimes}: the region of lower viscosity
right in front of the finger tip facilitates the advance of
the interface in the center of the channel. It is thus not
surprising that the lowest values of the finger width are
reached for $\We\sim 1$, where the variations of the viscosity
close to the tip are the strongest. Furthermore, this effect
increases with decreasing $\alpha$, as can be seen by comparing
the three curves obtained at \(\B=0.01\) in Fig.~\ref{fig_width_We_a}.
They coincide at small \(\We\) values since the first Newtonian
plateau is the same for all the curves. When $\We$ approaches
unity, the finger width decreases, with smaller $\alpha$ giving
rise to narrower fingers. This is to be expected since a smaller
$\alpha$ implies stronger variations of the viscosity with the
shear rate, and thus a stronger effective anisotropy. At
$\We\approx 10$, the curves cross. Now lower values of
$\alpha$ give rise to wider fingers. This is due to the
global decrease in viscosity in the shear-thinning fluid
already discussed above, together with the weakening of the
shear-thinning effect around the tip when the fluid enters the
second Newtonian plateau.

\begin{figure}[htbp]
\centering
\includegraphics[width=.8\textwidth]{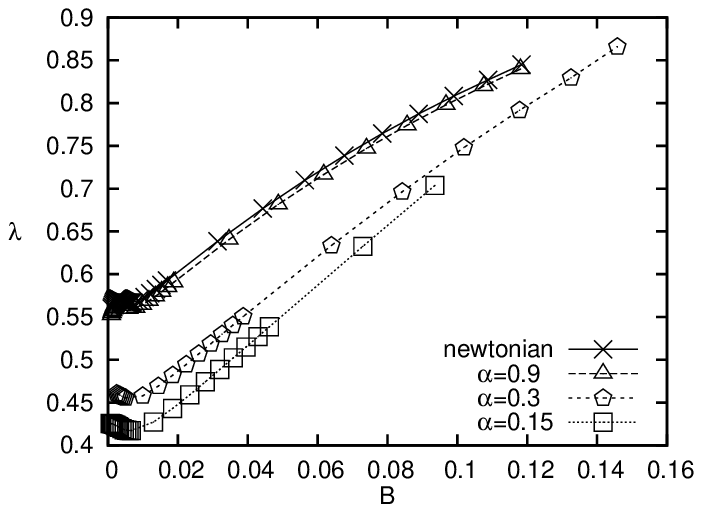}
\caption{Selected finger width as a function of $B$ for the 
effective viscosity of Eq.~(\protect\ref{2plateaux}) with three 
different values of $\alpha$,
and comparison to the Newtonian case; for all simulations,
\(\nu=5×10^{-3}\), \(\epsilon\)=0.02, $\Delta x=0.01$.}
\label{fig_comp_n_shth}
\end{figure}

From the preceding discussion, it is clear that for the effective viscosity law 
given by Eq.~\ref{2plateaux} the strongest shear-thinning effect occurs for 
$\We\sim 1$. Therefore, next, we fix $\We=1$ and study the selected finger width
as a function of $\B$ for various values of $\alpha$. As discussed previously,
in order to display the results in a meaningful way, finger widths need
to be plotted as a function of $B$, which can be calculated
{\em a posteriori} using  Eq.~(\ref{eq_btipdef}). Figure
\ref{fig_comp_n_shth} displays three selection curves for
\(\alpha=0.9, 0.3, 0.15\) and
\(\nu=5×10^{-3}\), compared with
the corresponding Newtonian curve at 
\(\nu=5×10^{-3}\). The curve for $\alpha=0.9$ is very close to the Newtonian
one; with decreasing values of $\alpha$, the selected finger width
decreases at fixed $B$, which is consistent with the picture
of an effective anisotropy increasing with $\alpha$. It should
also be noted that, as for the Newtonian fluid, a ``bump'' occurs
in the selection curve due to discretization effects; however,
for strongly shear-thinning fluids there clearly exists a range
of $B$ for which the solution is not affected by numerical artifacts,
and for which stationary fingers display a  width \(\lambda<1/2\),
which would be impossible for a Newtonian fluid.

\subsubsection{Strongly shear-thinning fluid}

We now turn to the case in which the shear-thinning effect is strong,
that is, the infinite-shear viscosity can be neglected in front of the
zero-shear viscosity, $\mu_1^{\infty}\to 0$ or $\alpha\to 0$.
Then, the high-shear plateau disappears, and the two-plateau law of
Eq. (\ref{carreau2plateaux}) 
becomes a one-plateau Carreau law
(see e.g. \cite{Barnes})
\begin{equation}
\label{1plateau}
\mu_1(\tau_1\dot\gamma)=\mu_1^0(1+(\tau_1 \dot\gamma)^2)^{(n-1)/2}.
\end{equation} 
This expression has been shown to provide a good fit to the aqueous
solution of the polymer xanthane used in the experiments of
Ref.~\cite{Lindner00,Lindner02}. In our simulations, we now use $n=0.5$.

Even for this simplified law, again no analytical expression exists for the
corresponding effective viscosity $\mueffi(\We|\vec\nabla p|)$ to use in
Darcy's law. We have therefore tabulated the effective viscosity
for our numerical calculations, following the procedure of 
Appendix~\ref{sec_darcy}. However, in the limit of large shear rates,
$\tau_1 \dot\gamma\gg 1$, Eq.~(\ref{1plateau}) reduces to the 
Ostwald-de-Waehle power-law viscosity,
$\mu_1\sim (\tau_1\dot\gamma)^{n-1}$, and it is easy to show
that the effective viscosity asymptotically behaves as
\begin{equation}
\label{effpowerlaw}
\mueffi(\We|\vec\nabla p|) \sim  \left(\We|\vec\nabla p|\right)^{(n-1)/n}.
\end{equation}

\begin{figure}[htbp]
\centering
\includegraphics[width=.8\textwidth]{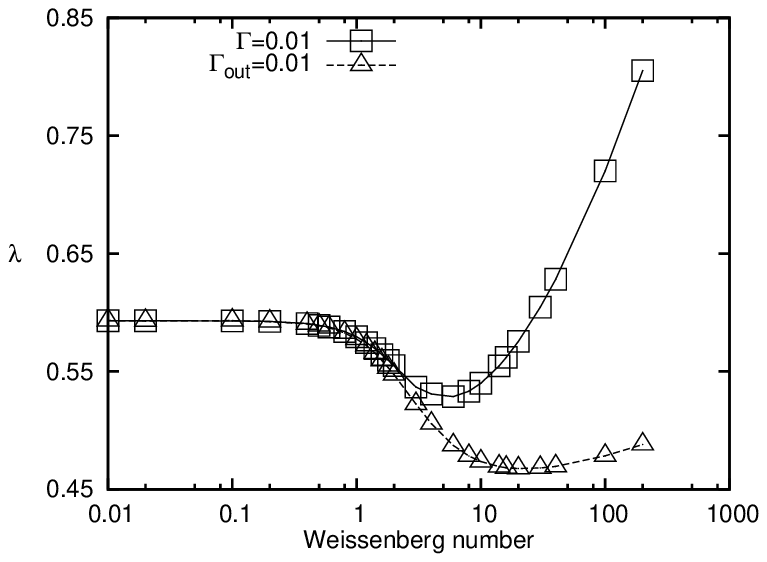}
\caption{Finger width versus $\We$ for the one-plateau law, either
at fixed $\B$, or at fixed $\B_{out}$, which is the dimensionless 
surface tension computed using the effective viscosity at the outlet 
[see Eq. (\ref{eq_Boutdef})].\label{fig_width_We_Bout}}
\end{figure}

Let us start by discussing the effect of the Weissenberg number. 
The curve of the finger width versus $\We$ obtained at constant 
$\B=0.01$ is shown in Fig. \ref{fig_width_We_Bout}. The onset of 
the shear-thinning regime occurs for $\We$ close to unity
as in the weakly shear-thinning case. The finger width then reaches a 
minimum once the whole tip region is in the shear-thinning regime.
When $\We$ is further increased, the width increases monotonously,
without exhibiting a plateau as in Fig. \ref{fig_width_We_a}. 
This is of course due to the fact that there is now no second plateau
in the viscosity law itself either. The viscosity continues to
exhibit a marked minimum at the tip, which implies that the
effective anisotropy is present for any $\We>1$. At the same
time, the viscosity everywhere in the shear-thinning fluid decreases
with increasing $\We$, which leads to an ever increasing value of
the tip selection parameter $B$, and therefore to an increase in
width.

Ideally, in order to separate the global variation of the viscosity
from the appearance of the effective anisotropy in the shear-thinning
fluid, the finger width should be studied at fixed $B$ instead of
fixed $\B$. This is difficult since $B$ can only be evaluated
{\em a posteriori}, as already discussed in Sec.~\ref{sec_params}.
However, a procedure can be devised that yields a clearer view:
instead of varying $\We$ at constant $\B$, one may also vary
simultaneously $\We$ and $\B$ to keep constant the dimensionless
surface tension defined with the viscosity {\em at the outlet}:
\begin{equation}
\B_{\rm out} = \frac{b^2\sigma}{12W^2 U_\infty
         \mu_1^0\mueffi_1(\We|\vec\nabla p_{\rm out}|)}
=\frac{\B}{\mueffi_1(\We|\vec\nabla p_{\rm out}|)}.
\label{eq_Boutdef}
\end{equation}
When $\We$ is varied, the new effective viscosity at the outlet
is computed through the pressure gradient value there,
which is the numerical solution of Eq.~(\ref{eq_pboundaryoutlet}).
$\B$ is then chosen to keep $\B_{\rm out}$ constant. 

The rationale for this procedure is the following: In the fully
shear-thinning regime, the whole tip is surrounded by a fluid 
region in which the effective viscosity scales as a simple power 
law. Intuitively, changing $\We$ in this regime should not alter 
the effective anisotropy at the tip, and the finger width selection 
should be governed by the only selection parameter left, the
dimensionless surface tension $B$. This scenario would be perfectly consistent
with the theoretical studies of Refs.~\cite{Corvera98,BenAmar99}.
In the power law regime, the viscosities at the tip and at the outlet
scale in the same way.
Therefore, for an inviscid pushing fluid, carrying out simulations 
with $\B_{\rm out}$ constant should leave the value of $B$ and hence the finger
width constant. Indeed, it can be seen in Fig.~\ref{fig_width_We_Bout} 
that the finger width for large Weissenberg numbers varies much less 
when keeping $\B_{\rm out}$ than when keeping $\B$ constant.
The residual increase of $\lambda$ with $\We$ is due to the finite
viscosity of the pushing fluid 2 ($\ratio>0$). 

Let us see the relation between $\B$, $B$, and $\B_{\rm out}$ in
detail:
In the power-law regime of Eq.~(\ref{effpowerlaw}),
$|\vec u|\sim |\vec\nabla p|^{1/n}$. Then, the viscosity of the Newtonian fluid is negelcted (that is,
we set $\ratio=0$), Eq.~(\ref{eq_Bgammarel}) becomes
\begin{equation}
B = \frac{U_\infty}{U_{\rm tip}}
\frac{\B}{\mueffi_1(\We|\vec\nabla p|_{\rm tip})}
  \sim \frac{U_\infty}{U_{\rm tip}^{n}}\B.
\end{equation}
Taking into account that, for a steady-state finger of width $\lambda$,
$U_{\rm tip}=U_\infty/\lambda$, and that $\lambda$ is a unique function
of $B$, we obtain
\begin{equation}
\frac{B}{\lambda(B)^{n}} \sim U_\infty^{1-n}\B \sim \We^{1-n}\B,
\end{equation}
and it is clear that $B$ (and hence $\lambda$) is fixed by the
product $\B\We^{1-n}$.

Similarly, it can be seen that $\B_{\rm out}$
defined by Eq.~(\ref{eq_Boutdef}) scales as $\sim \B\We^{1-n}$.
Therefore, for an inviscid fluid 2, keeping $\B_{\rm out}$ constant
amounts to keeping $B$, and hence the finger width, fixed.
The reason for the residual increase in $\lambda$  with increasing $\We$ 
but fixed $\B_{\rm out}$ in Fig.~\ref{fig_width_We_Bout} is the finite viscosity ratio 
$\ratio$. Indeed, this ratio, which is a constant independent
of $\We$, appears in the relation between $B$ and $\Gamma$, 
Eq.~(\ref{eq_Bgammarel}). Therefore, as the viscosity of the 
shear-thinning fluid 1 decreases with increasing $\We$, the
denominator gets smaller. As a result,  even at fixed $\B_{\rm out}$,  $B$ increases with increasing
$\We$,leading to an increase in the finger width $\lambda$.

This analysis shows that the dimensionless surface tension
$B$ in the power-law regime is determined by the parameter $\B\We^{1-n}$.
Our intuition that the effective anisotropy remains constant in that
regime suggests that the whole dynamics is controlled by this single
parameter. Substituting the expression for the power-law
effective viscosity, Eq. (\ref{effpowerlaw}) into Darcy's law,
Eq. (\ref{darcyless}), we get
\begin{equation}
\label{darcyforpowerlaw}
\vec u \sim -\left (\We|\vec\nabla p|\right )^{\frac{1-n}{n}} \; 
\left [\vec\nabla p + \B\kappa(\phi)\frac{\vec\nabla\phi}{2}\right ]
\end{equation}
 Assuming that we have a solution of the problem (i.e.: velocity field, pressure
 field and 
 finger shape, implicitly given by  $\phi$)  for a given set of $We$ and $\B$,
 we consider a situation where  the product $\B\We^{1-n}$ is kept constant while $\We$
 is multiplied by a positive value $\xi$ (this amounts to multiply $\B$ by
 $\xi^{n-1}$). In this case, considering eq.\ref{darcyforpowerlaw}, it is clear
 that if $\nabla p$ is also multiplied by $\xi^{n-1}$, the velocity field will be kept
 unchanged (and thus  obey the boundary conditions and the incompressibility
 condition for fluid 1).
 In the case where  fluid 2 is inviscid, its pressure gradient is zero, so the above
 rescaling for the pressure gradients in fluid 1 yields indeed a
 strictly valid solution in the whole domain and the dynamics depends only on 
 the reduced parameter $\B\We^{1-n}$.
 If the viscosity of fuid 2 is finite (but negligible), $0<\ratio<<1$,
 this rescaling is  a good approximation.

\begin{figure}[htbp]
\centering
\includegraphics[width=.8\textwidth]{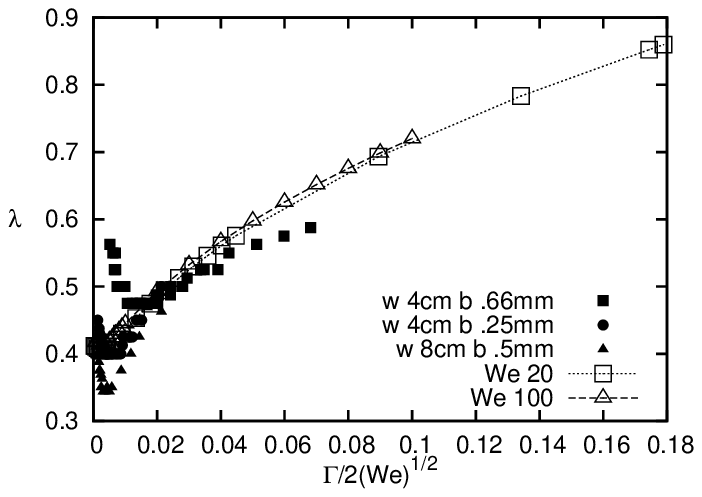}
\caption{Finger width as a function of the reduced parameter
$\We^{1-n}\B$ for two sets of simulations, and comparison to
the experimental data.}\label{fig_we_select}
\end{figure}

These predictions are indeed borne out by the simulation results.
In Fig.~\ref{fig_we_select}, we show the relative finger width
as a function of the only relevant parameter $\B\We^{1-n}$ (here,
$n=0.5$) for two series of simulations carried out at different values 
of $\We$. The two simulation curves, which differ by a factor 
of $5$ in the Weissenberg number, superimpose almost perfectly.

It turns out that the high-shear limit $\We|\vec\nabla p|\gg 1$ is 
also the relevant regime for the description of the experiments 
of Refs.~\cite{Lindner00,Lindner02}. Therefore, we show in the same
plot experimental data for various channel geometries. We have included 
complete data sets from Refs.~\cite{Lindner00,Lindner02}; these 
data exhibit first a decrease in the finger width with decreasing 
control parameter, but below a certain value they start 
to {\em increase} again, contrary to the theoretical predictions. 
This was attributed later \cite{Chevalier06} to the onset of inertial 
effects, which are obviously not contained in our model. Nevertheless, 
we have included all data points in our plot in order to
avoid an arbitrary cutoff.
The part of the data not affected by inertia (the part
with a positive slope) is quite close to our numerical curve.
It is interesting to note that to rescale the experimental
data, only the channel geometry and the flow rate $Q$ (or, 
equivalently, the finger speed and width) have to be known; 
no data on the viscosity of the tip are needed. Furthermore,
it is useful to stress that the scaling analysis makes it possible
to meaningfully compare simulations and experiments, even though
they are not carried out at the same parameters. The experimental
data correspond to high Weissenberg numbers ($\We\sim 10^3$)
and extremely small values of the viscosity ratio $\nu$ (since the pushing 
fluid is air); carrying out our simulations at these 
parameters would have been quite a numerical challenge.

\section{Conclusions}
\label{conclusions}

We have developed and validated a phase-field model for viscous
fingering in shear-thinning fluids in a Hele-Shaw cell. It can
be used for fluids with arbitrary shear-dependent viscosity,
provided that the viscosity function is not too steep 
to allow for the calculation of the effective viscosity function
by the method described in Appendix \ref{sec_darcy}. We have also shown that
the model is capable to describe the full crossover from Newtonian
to strongly shear-thinning behavior, and to make quantitative
contact with experimental results. It is therefore a useful and
robust tool for further investigations of the precise relationship
between the rheology of the shear-thinning fluid and the pattern
formation process.

We have investigated the selection of the finger width in the
channel geometry for two different shear-thinning laws. One is the model
fluid already used in Ref.~\cite{Fast01} that exhibits two
plateaux in the viscosity function at low and high shear rates,
and describes weakly shear-thinning fluids.
We have found that a narrowing of the fingers below the limit
$\lambda=1/2$ for Newtonian fluids is observed only in the
regime where most of the variations of the viscosity occur
in the vicinity of the tip. This confirms the idea that
the self-organization of the medium provides an effective
anisotropy leading to sharper finger tips. The second rheological law
investigated describes well the behaviour of the strongly shear-thinning
fluids used in the experiments of Refs.~\cite{Lindner00,Lindner02}.
Moreover, these exhibit a power-law viscosity at large shear rates. 
In this case, the system reaches a scaling regime  where 
the finger width depends  on a single parameter, 
 simply expressed in terms of the channel geometry and
the exponent of the viscosity law. This scaling makes it
possible to  compare simulations and experiments,
even though they are not carried out at the same parameters.
Reasonable agreement 
is obtained.

In the future, it would be interesting to use this model
for a systematic investigation of pattern selection as
a function of the viscosity law, especially in the regime
of narrow fingers. However, to attain this ``needle regime'',
improvements in the numerical algorithm will be needed, in
particular a refinement of the grid spacing at the interface.
This could be achieved using adaptive meshing algorithms.
Finally, the model can also be used without any difficulties
to simulate fingering in radial Hele-Shaw cells and to study
the transition from tip-splitting to stable dendritic growth.

\section*{Acknowledgements}
We thank A. Lidner for stimulating discussions.
R. F. acknowledges a Ramón y Cajal grant from Ministerio de Ciencia
e Innovación (MICINN, Spain), 
and further support from Universitat Rovira i Virgili under Project
No. 2006AIRE-01 and from MICINN under Projects No. 
CTQ2007-67435 and CTQ2008-06469/PPQ.

\appendix

\section{Darcy's law for non-Newtonian fluids}
\label{sec_darcy}
A (Newtonian) viscous fluid in a Hele-Shaw cell or porous
medium obeys Darcy's law: its velocity is proportional to the local
pressure gradient for not too high gradients, since then inertia can
be neglected. The proportionality constant can be understood as a
mobility, and it depends on the fluid viscosity and the
characteristics of the medium. In particular, in a Hele-Shaw cell
these ``medium'' characteristics are purely geometrical, since the
mobility appearing in Darcy's law is actually an average across
the cell gap. The underlying idea is to project
the actual three-dimensional problem into and effective bidimensional
problem in the plane of the glass plates, taking advantage of the
fact that the cell gap $b$ is much smaller than any other length
scale in the problem. In this projection procedure, one starts
from the Stokes equation for any fluid labelled by $i$,
\begin{equation}
\label{navierstokes}
\vec\nabla\cdot(\mu_i\vec\nabla\vec u)=\vec\nabla p.
\end{equation}
All quantities have their corresponding dimensions;
some of their dimensionless counterparts, as defined in particular by Eq. 
(\ref{adimumup}),
will only appear at the end of this Appendix
and will then be denoted by a tilde on top of their respective symbols.

In the left hand side of this Stokes Equation (\ref{navierstokes}),
$\partial_x$ and $\partial_y$ are neglected
with respect to $\partial_z$,
much stronger due to the small gap thickness, and one considers only
the in-plane flow ($x$ and $y$ directions).
We continue to denote the bidimensional versions by $\vec u$ and
$\vec\nabla$ to keep the notations simple. Note that, here, $\vec u$
Is a function of $z$. Integrating once, one gets
\begin{equation}
\label{2dnavierstokes}
\mu_i\partial_z\vec u=z\vec\nabla p.
\end{equation}

Darcy's law is then obtained by integrating once more to get
the in-plane velocity $\vec u$ and averaging the latter over the cell gap.
While this is straightforward for Newtonian fluids where the
viscosity is just a constant, in the non-Newtonian case where 
the viscosity depends on the shear $|\partial_z\vec u|$,
this is only possible if this function is invertible.
In the following, we detail the steps to obtain Darcy's law
in this case, in the spirit of Fast {\it et al.} \cite{Fast01}:

We rewrite the viscosity as
$\mu_i=\mu_i^0\tilde{\mu_i}(\tau_i^2|\partial_z\vec u|^2)$,
where $\mu_i^0$ is the zero-shear viscosity and
$\tilde{\mu_i}$ is a general, dimensionless viscosity function
of a dimensionless argument,
with $\tau_i$ some internal relaxation time of the fluid.
We take the modulus of Eq. (\ref{2dnavierstokes}) and multiply
it by $\tau_i$ to get
\begin{equation}
\label{mod2dnavierstokes}
\tilde{\mu_i}(s^2)s=\zeta z,
\end{equation}
where $s\equiv \tau_i|\partial_z\vec u|$ and
$\zeta\equiv\tau_i|\vec\nabla p|/\mu_i^0$.
As long as $s \tilde{\mu_i}(s^2)$ is an invertible function\footnote{In the case
of shear-thickening fluids, this condition is always verified, while in the case
of shear-thinning fluids, it yields the condition $2s^2\mu_i'+\mu_i >0$ which
writes $\mu_i/(2s^2)>-\mu'_i $ and can be interpreted as:
$\mu_i$ must not be too steep.},
this equation constitutes an implicit function
$s^2(\zeta^2z^2)$, which we reinject into $\tilde{\mu_i}(s^2)$ to
get $\tilde{\mu_i}(s^2(\zeta^2z^2))\equiv \mu_i^r(\zeta^2z^2)$.
We can now formally solve Equation (\ref{2dnavierstokes})
for $\partial_z\vec u$:
\begin{equation}
\partial_z\vec u = \frac{\vec\nabla p}{\mu_i^0\mu_i^r(\zeta^2z^2)} z
\end{equation}
and integrate it to get the in-plane velocity
\begin{equation}
\vec u = \frac{\vec\nabla p}{\mu_i^0} \int_{-b/2}^{z} 
\frac{z'dz'}{\mu_i^r(\zeta^2z^2)}
\end{equation}
Finally, we compute the gap-averaged velocity
\begin{equation}
\langle\vec u\rangle \equiv \frac{1}{b} \int_{-b/2}^{b/2} \vec u dz.
\end{equation}
After performing this latter integral by parts
and taking into account that the integrand is even, we obtain
\begin{equation}
\langle\vec u\rangle
= -2\frac{\vec\nabla p}{b\mu_i^0}\int_{0}^{b/2}\frac{z^2dz}{\mu_i'(\zeta^2z^2)}.
\end{equation}

At this point, we have obtained a relationship between the gap-averaged
velocity and the pressure gradient which is non-linear since the
pressure gradient appears not only in the prefactor, but also in
the integral (in the form of the factor $\zeta$). This relation
can then be used to define an effective viscosity that depends
on the pressure gradient. For computational purposes it is preferrable
to change the variable of integration from $z$ to $s$ according
to Eq.~(\ref{mod2dnavierstokes}).
In doing so, we go back from the inverse function $\mu_i'(\zeta^2z^2)$
to the original shear viscosity function $\tilde{\mu_i}(s^2)$.
We get that
\begin{eqnarray}
\int_{0}^{b/2}\frac{z^2dz}{\mu_i'(\zeta^2z^2)} = \frac{1}{\zeta^3} \int_{0}^\chi
\tilde{\mu_i}(s^2)s^2 \frac{d[\tilde{\mu_i}(s^2)s]}{ds}ds,
\\
{\rm where} \;\;\;\; \chi\equiv \frac{b\zeta}{2\mu_i'((b\zeta)^2/4)}
\;\;\;\; {\rm with} \;\;\;\; b\zeta=\frac{b\tau_i|\vec\nabla p|}{\mu_i^0}.
\end{eqnarray}
Integrating by parts once more 
we finally obtain
\begin{equation}
\langle\vec u\rangle
= -\frac{b^2\vec\nabla p}{\mu_i^0}(b\zeta)^{-
3}\left\{\tilde{\mu_i}^2(\chi^2)\chi^3-\int_0 ^\chi
\tilde{\mu_i}^2(s^2)s^2ds\right\}.
\end{equation}
This can be formally rewritten in the form of a Darcy's law,
\begin{eqnarray}
\langle\vec u\rangle = -\frac{b^2\vec\nabla p}{12\mu_i^0\mueffitil_i(b\zeta)},\\
{\rm where} \;\; \frac{1}{12\mueffitil_i(b\zeta)}\equiv
{b\zeta}^{-3}\left\{\tilde{\mu_i}^2(\chi^2)\chi^3-\int_0 ^\chi
\tilde{\mu_i}^2(s^2)s^2ds\right\},
\end{eqnarray}
with a mobility where the purely geometrical factor of 12
for Newtonian fluids has been replaced by a complicated function
of the variable $b\zeta$. This variable actually represents the
dimensionless shear.
Rewriting $\zeta$ in terms of the original quantities, and then
scaling the pressure as in the main text [Eq. (\ref{adimumup})],
it becomes
\begin{equation}
b\zeta = \frac{b\tau_i|\vec\nabla p|}{\mu_i^0} =
\frac{12\tau_i U_\infty}{b} \frac{\mu_1^0}{\mu_i^0} 
|\tilde{\vec\nabla}\tilde{p}|=
\left\{
\begin{array}{ll}
\quad \We|\tilde{\vec\nabla}\tilde{p}| &\;\;{\rm if}\;\;i=1 \\
(r/\ratio) \We|\tilde{\vec\nabla}\tilde{p}| &\;\;{\rm if}\;\;i=2
\end{array}
\right.,
\end{equation}
where the Weissenberg number $\We$ and the zero-shear viscosity ratio
$\ratio$ are defined by Eqs.~(\ref{globalwe}) and (\ref{viscratio})
in the main text, and
\begin{equation}
\label{tauratio}
r=\frac{\tau_2}{\tau_1},
\end{equation}
is the ratio of the characteristic time scales of the two fluids.
Note that we have here allowed for two different shear-dependent
viscosity laws. The global interpolated effective viscosity law
becomes then
\begin{eqnarray}
\nonumber
\muefftil(\phi)&=&\frac{1+\phi}{2}\mueffitil_1(\frac{12\tau_1 
U_\infty}{b}|\vec\nabla p|)
           +\frac{1-
\phi}{2}\ratio\mueffitil_2(\frac{12\tau_2|U_\infty}{b}\frac{\mu_1^0}{\mu_2^0}|\v
ec\nabla p|) \\
		     &=&\frac{1+\phi}{2}\mueffitil_1(\We |\tilde{\vec\nabla}\tilde 
p|)
           +\frac{1-\phi}{2}\ratio\mueffitil_2(\We 
|\tilde{\vec\nabla}\tilde p|r/\ratio).
\end{eqnarray}
The formulas of the main text, valid if fluid 2 is Newtonian, can then
be obtained by setting $r=0$ and $\mueffitil_2\equiv 1$.

\section{Numerical method}
\label{sec_numerics}
Here, we give some additional details about our numerical procedures.
Before performing the discretization of the dimensionless 
Eqs.~(\ref{darcyless}), (\ref{pfeq}), (\ref{eq_pboundaryoutlet}) 
and (\ref{inletpressure}) we choose to place ourselves in the 
frame moving at the velocity $U_\infty$ of the fluid at the outlet.
Using dimensionless units, the velocity field in this frame is \(\vec{v}\)
so that in the laboratory frame \(\vec{u}=\hat{y}+\vec{v}\),
(one should note that for a planar interface \(\vec{v}=0\)). Moreover,
we have chosen to solve the pressure equation considering a perturbation of
the average pressure gradient. This amounts to preconditionning the Poisson
operator which is ill-conditionned because of the presence of high viscosity
contrasts.

The spatial discretization of equations \eqref{darcyless}, 
\eqref{pfeq}, \eqref{eq_pboundaryoutlet} and \eqref{inletpressure} 
is based on finite differences on a staggered grid. The pressure 
and the phase field are evaluated at the mesh nodes, whereas 
the horizontal velocity component are evaluated at the mid-points of 
the horizontal links, and the vertical component at the mid-points 
of the vertical links. This staggering allows for a scheme 
that exactly guarantees mass conservation (\(\Div {\vv}=0\)). 
Time-stepping appears only in the phase field evolution equation. 
An explicit formulation of the time derivatives is retained,
which results in the following sequence (the time step is 
indicated as a superscript):
\begin{enumerate}
\item{} Given $\phi^{n}$ and \(p^{n-1}\), we calculate $\zet^{n}(\phi^{n},p^{n-
1})$ and $\kappa^{n}$.
\item{} The pressure field $p^{n}$ is given by the solution of the equation
\begin{equation}
\Div\left(\frac{\Nabla p^{n}}{\zet^{n}}\right) 
=\B\Div\left(\frac{\kappa^{n}\Nabla\phi^{n}}{\zet^{n}}\right)%
+\Nabla\left(\frac{1}{\zet^{n}}\right){\boldsymbol\cdot}\vec{y}.
\end{equation}
For a stationnary state, the effective viscosity and the pressure are 
the solutions of a fixed point
problem which converges in practice (for small enough time steps and suitable 
physical parameters).

\item{} The velocity is then directly evaluated by
\begin{equation}
{\vv}^{n} = -\frac{1}{\zet^{n}}\left\{\Nabla p^{n} - B 
\:\kappa^{n}\:\Nabla\phi^{n} + (\zet^{n}-1)\vec{y}\right\}.
\end{equation}
\item{} The phase field is timestepped,
\begin{equation}
\phi^{n+1}=\phi^{n}+{\rm dt}\left\{-{\vv}^{n}\cdot\Nabla\phi^{n}+\varepsilon^{-
2}(\phi-\phi^{3})^{n} + \nabla^2\phi^{n}
+ \kappa^{n}|\Nabla\phi^{n}|\right\}
\end{equation}
\end{enumerate}
Here, \(\zet\) is the dimensionless interpolated viscosity averaged 
over the gap in the most general case of appendix \ref{sec_darcy}.
Typically in our simulations, for both Newtonian and non-Newtonian
fluids the time step was of the order \(10^{-5}\).

The pressure is obtained by solving the linear system resulting from
the spatial discretization of the Poisson equation and associated
boundary conditions using
The Gauss-Seidel SOR method.  
The SOR solver is initialized with the pressure field of the preceding time 
step,
which helps to reduce significantly the amount of iterations needed
to achieve convergence of the pressure field after the few initial 
time steps. In all our simulations, the relaxation parameter 
\(\omega\) is set to 1,83. This value is chosen by trial and error in order
to limit the amount of overall iterations at each time step and 
to allow for fast enough parametric studies. The convergence 
criterion \(\varepsilon_{sor}\) was chosen so that the residual 
of the linear system was very close 
\footnote{To limit the number of operations per time step, the 
convergence criterion does not apply to the actual residual but 
to an intermediate computational vector, see numerical recipes. 
The actual residual was checked a posteriori to be close enough 
to this convergence criterion.} to \(10^{-5}\). A drawback of 
the SOR method is its sensitivity to the number of mesh points 
compared to more sophisticated methods such as, for instance, ADI.
Computations performed on a 2D test-case showed that the number of
iterations required to solve the problem on a square mesh with an
imposed accuracy of \(10^{-5}\) increases roughly as the number of
nodes in one direction when an relaxed ADI method necessitates a
fixed number of iterations once the problem is well resolved spatially.
However, for the grid resolution used in our parametric studies, 
\(N_x×{N_y}=100×500\), the number of iterations varies 
between 20 and 50 after the few initial time steps. Not surprisingly,
the number of iterations is found to depend on the 
choice of the physical parameters: more strongly shear-thinning 
fluids (higher \(n\) in Carreau's law) and higher $\We$ require
more iterations to converge. I is also found that varying the 
viscosity ratio can change the number of iterations needed to 
converge. Unexpectedly, when \(\ratio\) tends to zero the number 
of iterations per time step decreases, whereas the problem becomes 
numerically more challenging. This does not indicate that the 
SOR method works better for harder problems but that instead of 
solving the pressure field in front of the finger the iterations 
are used to solve the field in the finger which bears almost no 
information. The number of digits available in the numerical 
solution for the diffusion field is greatly influenced by 
the value of \(\ratio\). Since the overall available information 
in the numerical solution is set by \(\varepsilon_{sor}\) we 
adapted its value to roughly maintain constant the number of SOR 
iterations. In the Newtonian case, we found that to 
keep the SOR iterations value approximatively to 20 for \(\ratio\) 
values of 0.05 and 0.005, \(\varepsilon_{sor}\) needs to be set 
to \(10^{-4}\) and \(10^{-5}\), respectively.

\begin{figure}[htbp]
\centering
\includegraphics[width=.4\textwidth]{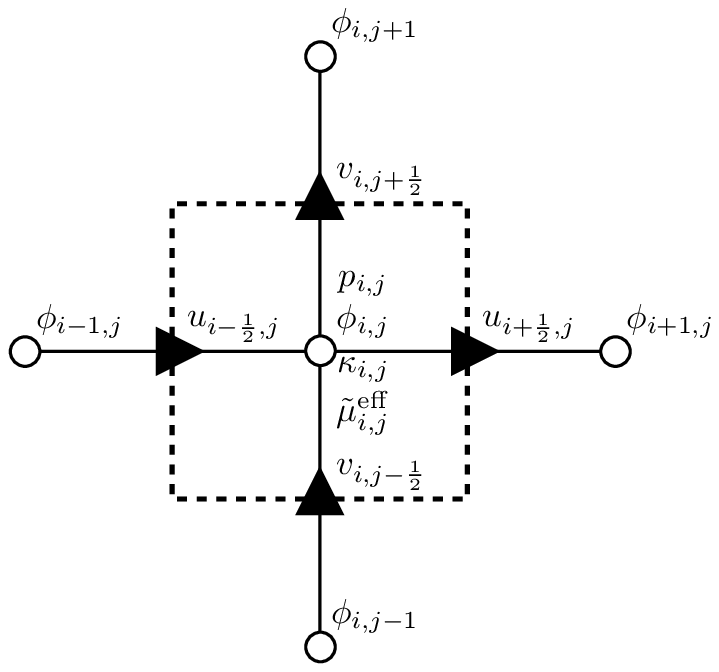}
\caption{Staggering of the fields}\label{fig_staggering}
\end{figure}

We conclude by a few remarks on the discretization.
In order to statisfy the incompressibility constraint, the Poisson equation
should be carefully discretized. Since the divergence operator is 
applied to equation \eqref{darcyless}, all the terms of
this equation should be evaluated in the middle of the links
surrounding the point where the discrete divergence applies (see 
fig.\ref{fig_staggering}). 
This amounts to:
\begin{align}
\Div\left(\frac{\Nabla p}{\zet}\right)_{i,j}&=&\left((\frac{\Nabla 
p}{\zet})_{i+\frac{1}{2},j}-(\frac{\Nabla
p}{\zet})_{i-\frac{1}{2},j}\right)\frac{1}{\Delta x}\nonumber\\
& &\mbox{}   +\left((\frac{\Nabla p}{\zet})_{i,j+\frac{1}{2}}-(\frac{\Nabla
p}{\zet})_{i,j-\frac{1}{2}}\right)\frac{1}{\Delta y},
\end{align}
with
\begin{equation}
\left(\frac{\Nabla
p}{\zet}\right)_{i+\frac{1}{2},j}=\frac{p_{i+1,j}-
p_{i,j}}{\frac{1}{2}(\zet{}_{i+1,j}+\zet{}_{i,j})\Delta x},
\end{equation}
for the left link, the discretization associated with the other links being 
staightforward. Similarly:
\begin{align}
\Div\left(\frac{\kappa\Nabla\phi}{\zet}\right)&=\left((\frac{\kappa\Nabla\phi}{\
zet})_{i+\frac{1}{2},j}-
(\frac{\kappa\Nabla\phi}{\zet})_{i-\frac{1}{2},j}\right)\frac{1}{\Delta 
x}\nonumber\\
&+\left((\frac{\kappa\Nabla\phi}{\zet})_{i,j+\frac{1}{2}}-
(\frac{\kappa\Nabla\phi}{\zet})_{i,j-\frac{1}{2}}\right)\frac{1}{\Delta y},
\end{align}
with,
\begin{equation}
(\frac{\kappa\Nabla\phi}{\zet})_{i+\frac{1}{2},j}=\frac{(\kappa_{i+1,j}+\kappa_{
i,j})(\phi_{i+1,j}-\phi_{i,j})}
{(\zet{}_{i+1,j}+\zet{}_{i,j})\Delta x},
\end{equation}
and
\begin{align}
\Nabla\left(\frac{1}{\zet}\right)_{i,j}{\boldsymbol\cdot}\vec{y}&=
\left(\frac{1}{\zet{}_{i,j+1}}-\frac{1}{\zet{}_{i,j-1}}\right)\frac{1}{\Delta 
x}.
\end{align}

In order to minimize rounding errors and ensure mass conservation, 
the spatial discretization of the velocity equation must be 
consistent with that of the Poisson equation for the pressure. 
Staggering results in shifting the velocity components in the 
direction to which they relate. This gives:
\begin{align}
v^x_{i+\frac{1}{2},j}&=-(\frac{\Nabla p}{\zet})_{i+\frac{1}{2},j}+B 
(\frac{\kappa\Nabla\phi}{\zet})_{i+\frac{1}{2},j},\\
v^y_{i,j+\frac{1}{2}}&=-(\frac{\Nabla p}{\zet})_{i,j+\frac{1}{2}}+B 
(\frac{\kappa\Nabla\phi}{\zet})_{i,j+\frac{1}{2}}+
\frac{1}{2}\left(\frac{1}{\zet{}_{i,j+1}}+\frac{1}{\zet{}_{i,j}}\right)-1.
\end{align}

The treatment of the phase field equation is straightforward and
analogous to that found in Ref.~\cite{Folch99b} except
for the advective term. The velocity on the the nodes is recovered
by a linear interpolation:
\begin{align}
({\vv}\cdot\Nabla\phi)_{i,j}&
 =\frac{1}{2}(v^x_{i+\frac{1}{2},j}+v^x_{i-\frac{1}{2},j})
  \frac{\phi_{i+1,j}-\phi_{i-1,j}}{\Delta x}\nonumber\\
&+\frac{1}{2}(v^y_{i,j+\frac{1}{2}}+v^y_{i,j-\frac{1}{2}})
  \frac{\phi_{i,j+1}-\phi_{i,j-1}}{\Delta y}.
\end{align}
The use of the phase field to compute a continuous equivalent of 
the curvature \(\kappa\) requires to introduce a numerical cutoff 
to avoid infinite values at centers of curvature. Wherever \(|\Nabla\phi|\)
is smaller than \(10^{-4}\), \(\kappa\) is replaced by 0.

\end{document}